\documentclass{aa}
\usepackage[varg]{txfonts}
\usepackage{graphicx}
\usepackage{newtxtext,newtxmath}
\usepackage{subcaption}
\usepackage[dvipsnames]{xcolor}
\RequirePackage[colorlinks=true,linkcolor=blue,citecolor=blue,urlcolor=blue]{hyperref}

\newcommand{\mvir}{{\rm M}_{\rm 500c}}
\newcommand{\rvir}{r_{\rm 500c}}
\newcommand{\mtc}{{\rm M}_{\rm 200c}}
\newcommand{\rtc}{r_{\rm 200c}}
\newcommand{\msun}{M_\odot}

\begin{document}

\title{The heart of galaxy clusters: demographics and physical properties of cool-core and non-cool-core halos in the TNG-Cluster simulation}
\titlerunning{Cool-Cores in TNG-Cluster}

\author{Katrin Lehle\inst{1}\thanks{E-mail: k.lehle@stud.uni-heidelberg.de}
\and Dylan Nelson\inst{1}
\and Annalisa Pillepich\inst{2}
\and Nhut Truong\inst{3}
\and Eric Rohr\inst{2}
} 

\institute{Universit\"{a}t Heidelberg, Zentrum f\"{u}r Astronomie, ITA, Albert-Ueberle-Str. 2, 69120 Heidelberg, Germany \label{1}
\and Max-Planck-Institut f\"{u}r Astronomie, K\"{o}nigstuhl 17, 69117 Heidelberg, Germany \label{2}
\and NASA/Goddard Space Flight Center, Greenbelt, MD 20771, USA \label{3}
}

\date{}

\abstract{
We analyze the physical properties of the gaseous intracluster medium (ICM) at the center of massive galaxy clusters with TNG-Cluster, a new cosmological magnetohydrodynamical simulation. Our sample contains 352 simulated clusters spanning a halo mass range of $10^{14} < {\rm M}_{\rm 500c} / M_\odot < 2 \times 10^{15}$ at $z=0$. We focus on the proposed classification of clusters into cool-core (CC) and non-cool-core (NCC) populations, the $z=0$ distribution of cluster central ICM properties, and the redshift evolution of the CC cluster population. We analyze resolved structure and radial profiles of entropy, temperature, electron number density, and pressure. To distinguish between CC and NCC clusters, we consider several criteria: central cooling time, central entropy, central density, X-ray concentration parameter, and density profile slope. According to TNG-Cluster and with no a-priori cluster selection, the distributions of these properties are unimodal, whereby CCs and NCCs represent the two extremes. Across the entire TNG-Cluster sample at $z=0$ and based on central cooling time, the strong CC fraction is $f_{\rm SCC} = 24$\%, compared to $f_{\rm WCC} = 60$\% and $f_{\rm NCC} = 16$\% for weak and non-cool-cores, respectively. However, the fraction of CCs depends strongly on both halo mass and redshift, although the magnitude and even direction of the trends vary with definition. The abundant statistics of simulated high-mass clusters in TNG-Cluster enables us to match observational samples and make a comparison with data. The CC fractions from $z=0$ to $z=2$ are in broad agreement with observations, as are radial profiles of thermodynamical quantities, globally as well as split for CC versus NCC halos. TNG-Cluster can therefore be used as a laboratory to study the evolution and transformations of cluster cores due to mergers, AGN feedback, and other physical processes.
}


\maketitle


\section{Introduction}

Within the Universe, galaxy clusters represent the most massive virialized objects. They originate from a hierarchical assembly process driven by the merging of smaller substructures. Although dominated in mass by dark matter, the main baryonic constituent of a cluster is the gas of the intracluster medium (ICM). The ICM is a hot, ionized plasma that permeates the cluster volume and can reach temperatures of up to $10^{7-8}$\,K \citep{mushotzky1978, mohr1999}. Its thermodynamical structure and evolution is subject to the impact of many intertwined physical processes \citep{fabian1994, cavaliere2002}.

On one hand, the gas of the ICM is heated by mechanisms including hydrodynamical shocks \citep{bourdin2013}, feedback from supermassive black holes (SMBHs) i.e. active galactic nuclei \citep[AGN;][]{ruppin2022, mittal2009}, and mergers \citep{mccarthy2007}. The ICM is also supported by non-thermal components including bulk motion \citep{liu2016, tamura2011} and turbulence \citep{mohapatra2019, banerjee2014}, magnetic fields \citep{taylor1993, soker1990}, and cosmic rays \citep{sanders2007, ruszkowski2023}. The thermal energy content of the ICM is lost due to radiative cooling, primarily via free-free bremsstrahlung \citep{lea1973}. However, this cooling is only effective when the gas is sufficiently dense, which is true in the cores of some, but not all, clusters. In particular, X-ray observations of the ICM reveal that the central cooling time can be short compared to the Hubble time \citep{cowie1977, fabian1977}. This implies that the ICM can rapidly cool down and form a so-called `cool core', or a cool-core (CC) cluster. On the other hand, some clusters do not appear to have such cool cores, and these are labeled as non-cool-core (NCC) clusters \citep{molendi2001}.

Early observational work found clear bimodalities in the properties of cool-core versus non-cool-core clusters. For example, \cite{hudson2007} identify a bimodality in the distribution of temperature gradients, splitting the clusters in the HIFLUGCS sample roughly in half. \cite{cavagnolo2009} find a distinct gap in the central entropy excess distribution, splitting the sample also roughly into two equal halves \citep[other studies find no clear bimodalities, e.g.][]{croston2008, pratt2010}.

Different physical properties of cluster cores may be more or less linked to cool-core state, and/or more or less useful in classifying different types of clusters. \cite{hudson2010} consider which cluster core property is best suited to distinguish between two, or perhaps three, distinct classes of clusters. They find that the central cooling time shows high significance of bimodality and is best suited as a CC metric for low-redshift clusters with high-quality data available. For high-redshift clusters, the cuspiness can best distinguish between CCs and NCCs. With the X-ray flux-limited \textsc{HIFLUGCS} clusters, they find that the various CC criteria lead to different CC fractions in the sample. 

The overall fraction of cool-core clusters, as a function of mass, redshift, and halo/galaxy properties, directly reflects the complex baryonic physics of cluster assembly. Unfortunately, observational inferences of cool-core fractions are difficult, as they strongly depend on the defining CC metric \citep{hudson2010}, as well as the cluster sample selection \citep{andrade-santos2017}. Most problematic, cool-core clusters, being characterized by pronounced peaks in their X-ray surface brightness and higher X-ray luminosity at given mass, are more easily detected in flux-limited X-ray surveys. This detection preference, known as the `CC bias' \citep{eckert2011}, leads to a probable overestimation of the cool-core fraction within X-ray samples.

Recently, Sunyaev Zel’dovich (SZ) surveys have begun to enable mass-limited cluster samples. With this technique, \cite{andrade-santos2017} conclude that CC fractions from X-ray selected samples are 2.1 - 2.7 times larger than in SZ-selected samples, depending on CC criterion. Using the 164 SZ-selected clusters from the \textit{Planck} ESZ sample, they present distributions of cluster properties that have no bimodal characteristics. This finding agrees with other surveys finding unimodal distributions of central cluster properties, both SZ selected \citep{planckcollaboration2011, mcdonald2013, rossetti2017} and Dark Energy Survey optically selected \citep{graham2023}.

Beyond $z=0$, the redshift evolution of clusters directly constrains their assembly histories and growth. Perhaps surprisingly, the cores of observed clusters show no significant evolution since $z\sim 2$ \citep{mcdonald2013, sanders2018, ruppin2021}. In particular, while the bulk of the ICM evolves in a self-similar manner, cores do not and are consistent with no redshift evolution \citep{mcdonald2017}. This implies that cool cores, if present, form early and do not significantly change in e.g. size or structure. As a result, a stable and persistent heating source(s) must offset otherwise rapid cooling and enable clusters to grow cool cores \citep{mcdonald2013}. Similar findings hold for the evolution of clusters that resemble the progenitors of well-studied nearby clusters \citep{ruppin2021}.

Modeling the complex physics in high-mass galaxy clusters is a challenging regime for numerical simulations. In addition, comparisons with cluster observations require detailed forward modeling and consideration of the systematics and biases at play. Early cosmological simulations incorporating gas hydrodynamics could directly assess cluster CC status \citep{burns2007, planelles2009a}. More recently, cosmological simulations such as IllustrisTNG have begun to study cool-core populations with broadly realistic baryonic feedback models \citep{barnes2018}. The largest volume simulations, including Magneticum \citep{dolag2015}, MilleniumTNG \citep{pakmor2023}, FLAMINGO \citep{schaye2023a}, extend (N)CC modeling the most massive clusters. Simultaneously, zoom simulation projects such as RHAPSODY-G \citep{hahn2017a}, DIANOGA \cite{rasia2015}, The Three Hundred \citep{cui2018}, and C-EAGLE/Hydrangea \citep{barnes2017a,bahe2017} can also capture (N)CC populations.

Two key benchmarks for simulated clusters are the distributions of central cluster properties, and the cool-core fraction. Some cosmological simulations produce continuous CC distributions \citep{barnes2018, kay2007}, while others identify bimodalities \citep{hahn2017a}. The inferred cool-core fractions vary strongly between simulations. For example, \cite{burns2007} and \cite{planelles2009a} find CC fractions of $\sim 16\%$, while \cite{hahn2017a} obtain a CC fraction of 44\%, versus 38\% for \cite{rasia2015}. Employing the same TNG galaxy formation model, \cite{barnes2018} infer cool-core fractions of 1-21\% in TNG300, depending on CC criterion, while MTNG finds $f_{\rm CC} \simeq 8$\% \citep{pakmor2023}. Overall, the comparison of CC fractions with observations, and among simulations, is complicated by the diversity of CC criteria in use.

The physical drivers of cool-core clusters, and the transformation(s) between CCs and NCCs, are not well understood. Early mergers may be responsible for destroying CCs and thereby producing NCCs \citep{burns2007}, while late mergers are not capable of destroying a CC \citep{poole2008}. However, more recent simulations find that late mergers are able to destroy cool cores \citep{rasia2015}. \cite{barnes2018} conclude that the fraction of relaxed clusters is similar for cool-core and non-cool-core populations, implying that mergers alone are not responsible. They may be necessary but not sufficient for destroying CCs, as their efficacy depends on the amount of angular momentum in the merger \citep{hahn2017a}.

AGN feedback is an additional mechanism that could change the core state of clusters. It is a major heating source of the ICM, that can stop cooling flows and preserve the cool-core structure \citep{li2015, rasia2015, gaspari2013, gaspari2012}. However, it is unclear if AGN feedback significantly impacts cluster core status. Idealized simulations give conflicting results: \cite{guo2009} and \cite{barai2016} find that AGNs can transform CCs to NCCs, while \cite{ehlert2023} find that AGN feedback of light jets cannot.

In this study we investigate the galaxy cluster population of the new TNG-Cluster simulation. This is a suite of high-resolution ($m_{\rm gas} \simeq 10^7 M_\odot$) zoom simulations of massive galaxy clusters (M$_{\rm{halo}}\sim 10^{15} M_\odot$) employing the IllustrisTNG galaxy formation model. TNG-Cluster is well suited to study the CC population, as the simulation offers a unique combination of high-mass galaxy clusters and high resolution, combined with a comprehensive physical model including SMBH feedback. The goals of this work are: (i) to provide a census of the cool-core cluster population and CC fractions of TNG-Cluster at $z=0$; and (ii) to study and quantify TNG-Cluster predictions for the redshift evolution of central ICM properties and CC fractions of the (N)CC populations.

This is one of a number of papers where we highlight first science results from TNG-Cluster. Following an overview of the simulation suite and the basic properties of clusters (\textcolor{blue}{Nelson et al. submitted}), these are: a census of gas motions and kinematics from cluster cores to outskirts (\textcolor{blue}{Ayromlou et al. submitted}), the inference of cluster kinematics from high-resolution X-ray spectroscopic data (e.g. Hitomi/XRISM/LEM; \textcolor{blue}{Truong et al. submitted}), an identification of merging clusters producing a diversity of radio emission and radio relic features (\textcolor{blue}{Lee et al. submitted}), and the retention and observability of the circumgalactic medium of cluster satellite galaxies (\textcolor{blue}{Rohr et al. submitted}).

This paper is organized as follows. Section \ref{sec_methods} describes the TNG-Cluster simulation, while Sec.~\ref{ch:CCcritDef} defines the six criteria that we use to categorize clusters as (N)CCs. In Sec.~\ref{sec_z0} we present the (N)CC population at $z=0$, focusing on thermodynamical properties (Sec.~\ref{subsec_thermoProp}), the distribution of core properties (Sec.~\ref{subsec_CCvsNCC}), their mass dependence (Sec.~\ref{subsec_CCcritVsM500}), the CC fraction (Sec.~\ref{subsec_fcc}), and ICM internal structure (Sec.~\ref{subsec_profilesCCvsNCC}). We then study redshift evolution in Sec.~\ref{sec_zevo}: of central physical properties (Sec.~\ref{ch:CCcritvsZ}), CC fractions (Sec.~\ref{subsec_fccVSz}), and cluster profiles, split by mass and (N)CC status (Sec.~\ref{subsec_profilesVSz}). In Sec.~\ref{sec_conclusions} we summarize our findings.


\section{Methods} \label{sec_methods}

\subsection{The TNG-Cluster Simulation}

TNG-Cluster is a collection of 352 high-resolution zoom simulations targeted to study massive galaxy clusters.\footnote{\url{www.tng-project.org/cluster}} This project is a spin-off of the IllustrisTNG project \citep[hereafter TNG;][]{nelson2018, pillepich2018a, marinacci2018, springel2018, naiman2018}, a suite of cosmological gravo-magnetohydrodynamical simulations of galaxy formation and evolution. The original TNG suite simulated three different volumes: TNG100 and TNG300, with larger volumes of box length $\sim 100$\,cMpc and $\sim 300$\,cMpc, respectively, and TNG50, a smaller, higher-resolution box with side length $\sim 50$\,cMpc \citep{pillepich2019, nelson2019}. 

The TNG simulations use the AREPO code \citep{springel2010} and solve the coupled self-gravity and ideal magnetohydrodynamics (MHD) equations \citep{pakmor2013, pakmor2011}. A notable aspect of the TNG simulation is its comprehensive and thoroughly validated physical model for galaxy formation and evolution, described in detail in \cite{weinberger2017} and \cite{pillepich2018}. TNG-Cluster employs this same, unchanged model and hence includes the key processes relevant to the formation and evolution of galaxies and galaxy clusters, including heating and cooling of gas, star formation, evolution of stellar populations and chemical enrichment, stellar feedback, as well as growth, merging and multi-mode feedback from SMBHs. TNG-Cluster adopts the fiducial TNG cosmology, consistent with the \cite{planckcollaboration2016}: $\Omega_m = 0.3089$, $\Omega_b = 0.0486$, $\Omega_\Lambda = 0.6911$, $H_0 = 100 h$\,km s$^{-1}$Mpc$^{-1}$ = 67.74\,km\,s$^{-1}$Mpc$^{-1}$, $\sigma_8 = 0.8159$, and $n_s = 0.9667$.

The TNG-Cluster suite is an extension of the TNG300 simulation as it improves upon its sampling and statistics of halos at the high-mass end. The target clusters for the zoom (re-)simulations were selected from a large dark matter only run of a periodic box of volume (1\,Gpc)$^3$. Halos are chosen solely based on halo mass at $z=0$, such that all halos with $\log_{10} (\mtc/\msun) > 15.0 $\footnote{$\mtc$ is the mass enclosed within $\rtc$, which in turn is the radius enclosing a sphere with average density 200 times denser than the critical density of the universe at a given redshift. $\rvir$ is the radius enclosing a sphere with average density 500 times denser than the critical density of the universe at a given redshift, and $\mvir$ is the mass enclosed in $\rvir$.} were included, while for masses $14.5 < \log_{10} (\mtc/\msun) < 15.0$ halos were randomly selected in 0.1 dex mass bins to compensate for the drop-off in halo mass of the TNG300 sample and to achieve a uniform distribution -- see Figure 1 in \textcolor{blue}{Nelson et al. (submitted)}. The TNG-Cluster simulation has the same resolution as TNG300-1, i.e. m$_{\rm{gas}} = 1.2\times10^7\msun$ and m$_{\rm{DM}} = 6.1 \times 10^7\msun$. 

Halos are identified using the standard friends-of-friends (FoF) algorithm with a linking length of $b=0.2$. Substructures are identified using the \textsc{SUBFIND} routine \citep{springel2001} and are linked across different snapshots via the \textsc{SubLink} algorithm \citep{rodriguez-gomez2015}.

\subsection{Thermodynamical Quantities}\label{ch:thermDef}

For the computation of all thermodynamical properties of the simulated clusters, such as density, entropy, temperature, cooling time, and pressure, we consider all gas that is gravitationally bound to the central galaxy, i.e. to the cluster, according to \textsc{SubFind}. We exclude star forming gas cells since their temperature is set by an effective equation of state as per the two-phase interstellar medium model \citep{springel2003}. Unless otherwise stated, average quantities in map pixels or radial profile bins are mass-weighted means, i.e. weighted by the mass of each contributing gas cell.

\subsection{Cool-Core Criteria}\label{ch:CCcritDef}

{\renewcommand{\arraystretch}{1.3}
\begin{table*}
    \caption{Summary of the cool-core criteria considered throughout. Various previous works, both observational and theoretical, have used one or more central thermodynamical properties to separate strong cool-core (SCC), weak cool-core (WCC), and non-cool-core (NCC) clusters. Following \cite{barnes2018}, we consider six different criteria: $t_{\rm{cool,0}}$, $K_0$, $n_{\rm e,0}$, $\alpha$, $C_{\rm phys}$, and $C_{\rm scaled}$. In addition, the radius (aperture) within which each is measured typically varies, as indicated. Note that these radii and thresholds are adopted from previous studies, and do not necessarily produce an optimal identification of CC vs NCC clusters.}
  \begin{tabular}{llllll}
    \hline\hline
    Physical property & & Aperture & SCC threshold & WCC threshold & NCC threshold\\
    \hline
    Central cooling time & $t_{\rm{cool},0}$ & $0.012\, \rvir$ & $<1$\,Gyr& $1\, \rm{Gyr} \leq t_{\rm{cool},0} < 7.7\, \rm{Gyr}$ & $\geq7.7$\,Gyr \\
    Central entropy & $K_0$ & $0.012\, \rvir$&$\leq 22 $\,keV cm$^2$ & $22 < K_{\rm 0} / (\rm{keV\, cm}^2) \leq 150$& $> 150$\,keV cm$^2$ \\
    Central electron density & $n_{\rm{e},0}$ &$0.012\, \rvir$&$> 1.5 \cdot 10^{-2}$\,cm$^{-3}$ &$0.015 \geq n_{\rm e,0} / \rm{cm^{-3}} > 0.005$ & $\leq 0.5 \cdot 10^{-2}$\,cm$^{-3}$\\
    Cuspiness & $\alpha$ & $0.04\, \rvir$ & $>0.75$ & $0.75 \geq \alpha > 0.5$ & $\leq 0.5$ \\
    Physical concentration & $C_{\rm{phys}}$ & 40\,kpc, 400\,kpc & $>0.155$ & $0.155 \geq C_{\rm{phys}} > 0.075$ & $\leq 0.075$\\
    Scaled concentration & $C_{\rm{scaled}}$ & $0.15\, \rvir, \rvir$&$> 0.5$&  $0.5 \geq C_{\rm{scaled}} > 0.2$& $\leq 0.2$  \\
    \hline\hline
  \end{tabular}
  
  \label{tab:CCcuts}
\end{table*}}

Past studies have considered several different criteria to define cool-core (CC) versus non-cool-core (NCC) clusters. These criteria are motivated by theoretical considerations as well as observational findings. While theoretically motivated definitions may be less directly observable, they often reflect more closely the intrinsic physical properties of the systems or the results of numerical simulations. In particular, the ICM in clusters is observed at a variety of wavelengths, with the types of data varying from cluster to cluster and also depending on the redshift. The availability and quality of data may sometimes prevent the derivation of quantities necessary for the measurement of some criteria.

Following \cite{barnes2018}, in this work we examine six different CC definitions and describe them in detail below. Additionally, we summarize the thresholds for the employed CC metrics in Table~\ref{tab:CCcuts}. In all cases, when calculating the CC criteria, we aim to consider the gas that would be observed in a galaxy cluster, namely the hot gas cooling via Bremsstrahlung. To that end, as for the general thermodynamical properties above (Sec.~\ref{ch:thermDef}), we consider all gas cells that are gravitationally bound to the central galaxy, i.e. the cluster, according to \textsc{SubFind}. We further include only non-star-forming gas, cells that are cooling\footnote{We consider the net cooling rate, that is a direct output of the simulation, to define the cooling status of cells. If the net cooling rate is negative, the cells are cooling.} and have a temperature T>10$^6$K. We have checked and using all FoF cells instead of the gravitationally bound cells makes no difference for our results. As the halo center we choose the gravitational potential minimum of the cluster. This choice of halo center can considerably impact the CC/NCC classification and is typically different than what adopted in observational work. 

\subsubsection{Central cooling time}

In order for a cluster to form a cool core, the cooling time scale must be sufficiently short to allow gas to rapidly cool down. In the following, we consider the definition of core status based on the central cooling time as the most fundamental, and use it as our fiducial choice. The cooling time of the gas is computed as
\begin{equation}
    t_{\rm{cool}} = \frac{3}{2}\frac{(n_{\rm e}+n_i)k_{\rm B} T}{n_{ \rm e}n_i\Lambda} \, ,
\end{equation}
where $n_{\rm e}$ is the electron number density, $n_i$ the ion number density, $T$ the temperature, $\Lambda$ the cooling function, and $k_{\rm B}$ the Boltzmann constant. The cooling function $\Lambda$ is the instantaneous net cooling function of each gas cell, as output by the simulation itself. Consistent with previous work, we measure the cooling time within a 3D aperture of 0.012$\rvir$ \citep[e.g.][]{mcdonald2013,barnes2018}. 

A cluster is expected not to form a cool core, or a cooling flow, when its central cooling time is larger than the age of the Universe at the redshift of formation. 
However, to make the definition independent of redshift, it is typical to define clusters with $t_{\rm{cool},0}\geq 7.7$\,Gyr as NCCs \citep{mcdonald2013}. Clusters are defined as a SCC if the central cooling time is $<1$\,Gyr and as weak cool-core (WCC) if 1\,Gyr $\leq t_{\rm{cool, 0}} < $  7.7\,Gyr. 

The threshold of 7.7\,Gyr is a somewhat arbitrary choice. However, it arises from the idea that clusters at $z>1$ can be considered protoclusters. If we take $z=1$ as the average formation redshift of $z=0$ clusters, this corresponds to a lookback time of 7.7\,Gyr. Assuming clusters are born without a cool core, clusters with cooling times exceeding this limit have not had sufficient time to develop a cool core, motivating their classification as NCC. We return to the limitations of this choice below.

\subsubsection{Central entropy}

In absence of non-gravitational processes and in a global quasi-hydrostatic equilibrium, high-entropy gas is expected to rise in the cluster potential whereas low-entropy gas is expected to sink towards the center. Thus, cool-core clusters are clusters with low-entropy at their center. We compute the entropy of gas as
\begin{equation}
    K=k_{\rm B} Tn_{\rm e}^{-2/3}\, .
\end{equation}
We measure the entropy in the same 3D aperture of 0.012$\rvir$ and use the thresholds stated by \cite{hudson2010} to classify clusters: we define SCCs to have $K_0 = K(r < 0.012 \rvir) \leq 22 $\,keV cm$^2$. WCCs are clusters with 22\,keV\,cm$^2 < K_0 \leq 150$\,keV\,cm$^2$ and NCCs have $K_0 > 150$\,keV cm$^2$.

\subsubsection{Central electron number density}

To keep the gas in a cooling cluster core in pressure equilibrium and to balance the pressure loss due to the lower temperatures, gas must have high density. Thus, the central gas density can also be used to classify the cooling state of clusters. In practice, observations often consider the electron number density, as this quantity can be more directly inferred from data than e.g. cooling time \citep{barnes2018}. Following \cite{hudson2010}, we define clusters as SCC if the central electron number density  $n_{\rm{e}, 0} > 1.5 \cdot 10^{-2}$\,cm$^{-3}$. WCCs are clusters with $1.5 \cdot 10^{-2}$\,cm$^{-3} \geq n_{\rm{e}, 0} > 0.5 \cdot 10^{-2}$\,cm$^{-3}$ and NCCs have $n_{\rm{e}, 0} \leq 0.5 \cdot 10^{-2}$\,cm$^{-3}$.

Although motivated by observation considerations, here we measure the true intrinsic central density of the simulated clusters without replicating any observational procedure. Namely, to measure $n_{\rm e,0}$ from the simulation data, we take the mean density of gas within a 3D aperture of 0.012$\rvir$, considering only the cells with the specifications described above.

\subsubsection{Slope of the density profile - the cuspiness parameter}

Limited angular resolution makes the extraction of a temperature profile difficult for high redshift clusters. For that reason, \cite{vikhlinin2007} suggest a CC metric that is solely based on X-ray imaging data, leveraging the correlation between the presence of a cool core and the X-ray morphologies.

Clusters with a short central cooling time have been observed to exhibit central peaks in the X-ray distribution, while observed clusters without a cool core usually have flat cores. To capture this cusp/core structure, the cuspiness parameter is defined as
\begin{equation}
    \alpha = -\frac{\mathrm{d}\log n_e(r)}{\mathrm{d}\log r}\bigg |_{r=0.04\rvir} \, .
\end{equation}
For this analysis, we again extract an {\it intrinsic} cuspiness parameter, by computing the 3D radial profile of the electron number density with 50 logarithmically spaced bins from 10$^{-3}$ to $1.5 \rvir$. Following previous analyses \citep{vikhlinin2007, hudson2010, barnes2018}, we define SCCs to have $\alpha > 0.75$, WCCs to have $0.5 < \alpha \leq 0.75$, and NCCs to have $\alpha \leq 0.5 $.

\subsubsection{X-ray concentration parameter}

Motivated by the unavailability of spatially resolved data at high redshift, the X-ray concentration is also used to define CC status. To do so, the luminosity in the center is divided by the luminosity enclosed within a larger region,
\begin{equation}
    C_{\rm{phys}} = \frac{L_X^{0.5-5\rm{keV}}(r_{\rm p}<40\,\rm{kpc})}{L_X^{0.5-5\rm{keV}}(r_{\rm p}<400\,\rm{kpc})}\, .
\end{equation}
$L_X^{0.5-5\rm{keV}}$ is the X-ray luminosity in the 0.5-5 keV energy range and $r_p$ the projected radius. We compute the concentration parameter from 2D maps of the X-ray luminosity (see below). We define SCCs as clusters with $C_{\rm{phys}} > 0.155$, WCCs have $0.075 < C_{\rm{phys}} \leq 0.155$ and NCCs fulfil $C_{\rm{phys}}\leq 0.075$.
\begin{figure*}
    \centering
	\includegraphics[width=0.95\textwidth]{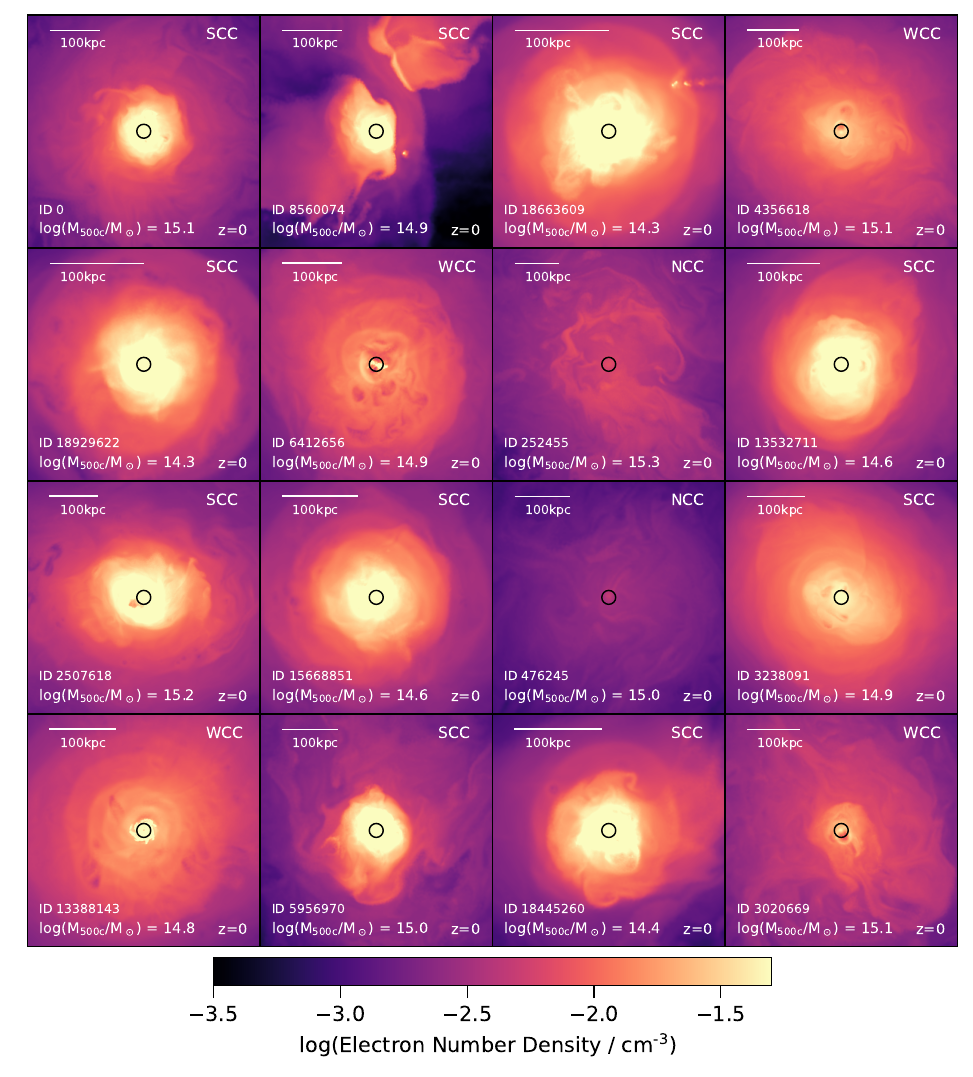}
    \caption{Visualization of the electron number density in the core region of 16 clusters of TNG-Cluster at $z=0$. Each panel shows the central 20\% of $\rvir$, projected along a line of sight of 15\,kpc. The halo mass $\mvir$ of each cluster is indicated in the panels. In the upper right of each panel we also label the core state of each cluster based on the central cooling time. The aperture of 0.012 $\rvir$ used to compute the central cooling time is depicted by black circles in the maps (see text). The clusters were chosen to reflect the diversity of the simulated cluster masses and to show the wide range of different morphologies of the gas.} 
    \label{fig:CoreGallery_ne} 
\end{figure*}

\begin{figure*}
    \centering
	\includegraphics[width=0.95\textwidth]{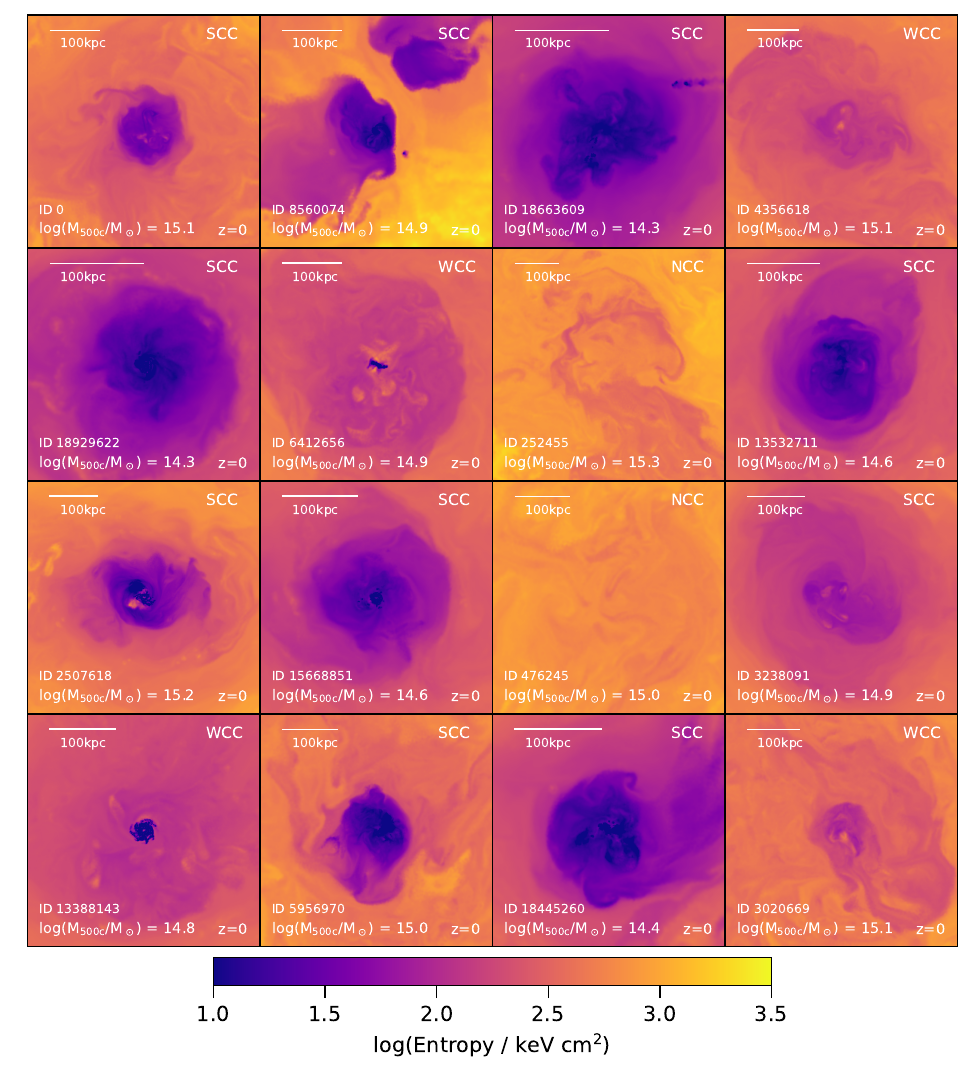}
    \caption{Visualization of the entropy distribution in the core region of the same 16 clusters as in Figure~\ref{fig:CoreGallery_ne} at $z=0$. In some of the clusters a distinct core with a lower entropy is visible. TNG-Cluster produces a diverse cluster population in terms of spatially resolved entropy morphologies, some of which resemble observed clusters.}
    \label{fig:CoreGallery_K}
\end{figure*}

\begin{figure*}
    \centering
	\includegraphics[width=0.95\textwidth]{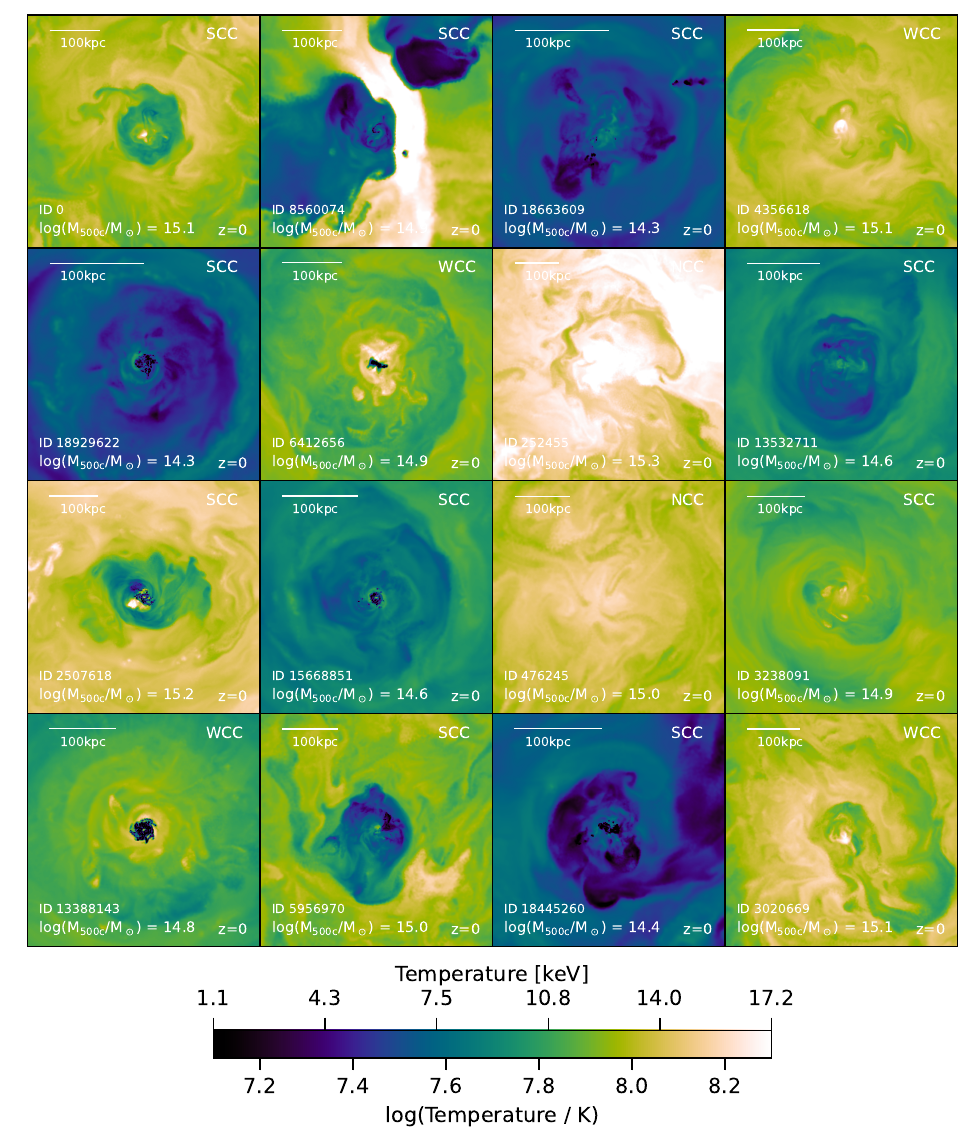}
    \caption{Visualization of the temperature distribution in the core region of the same 16 clusters as in Figure~\ref{fig:CoreGallery_ne} at $z=0$. The distinct cores of some clusters are also visible in the temperature maps. In contrast to the previous figure, the average temperature in the panels depends strongly on the mass of the cluster.}
    \label{fig:CoreGallery_T}
\end{figure*}

\begin{figure*}
    \centering
	\includegraphics[width=0.95\textwidth]{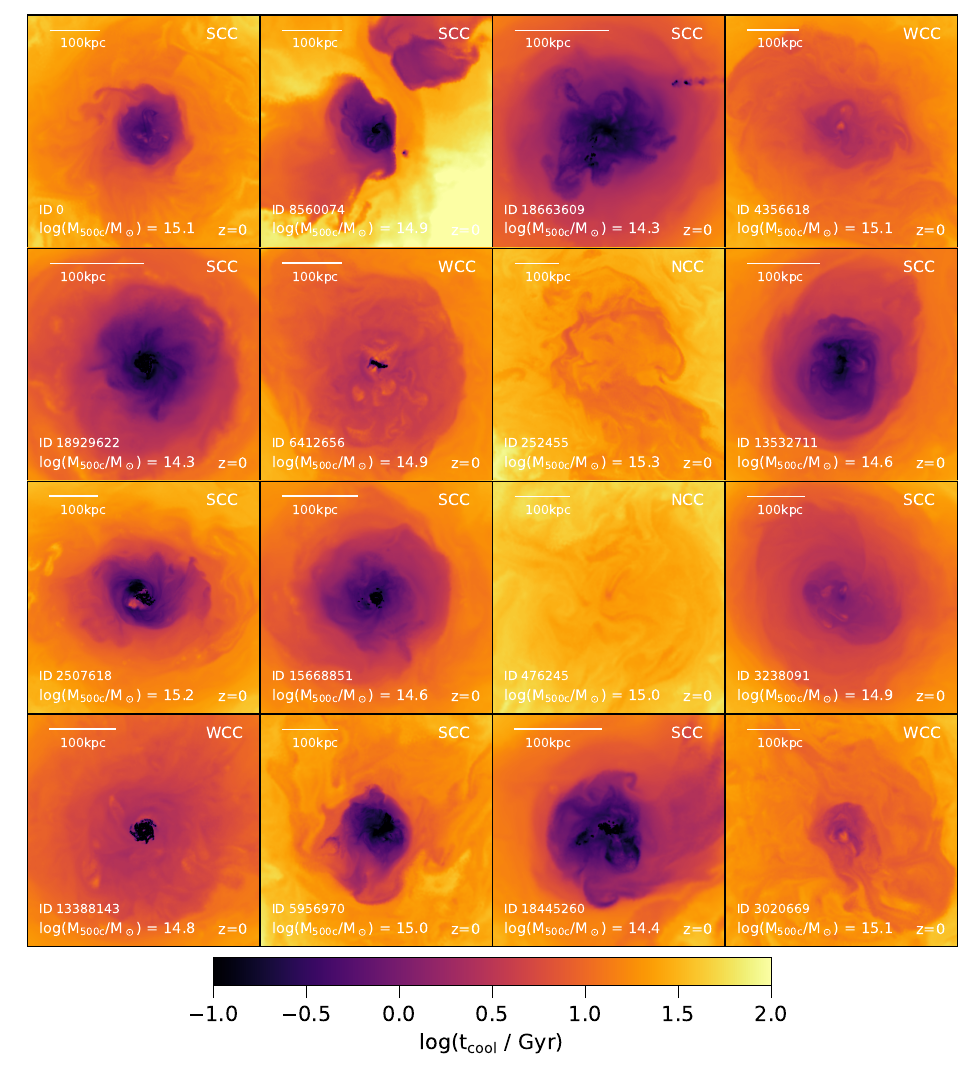}
    \caption{Visualization of the cooling time distribution in the core region of the same 16 clusters as in Figure~\ref{fig:CoreGallery_ne} at $z=0$. The SCCs exhibit distinct low cooling time cores.}
    \label{fig:CoreGallery_tcool}
\end{figure*}

The classification as well as the thresholds were originally introduced by \cite{santos2008}. \cite{maughan2012} also introduced a scaled version of the concentration parameter
\begin{equation}
    C_{\rm{scaled}} = \frac{L_X^{0.5-5\rm{keV}}(r_{\rm p}<0.15\rvir)}{L_X^{0.5-5\rm{keV}}(r_{\rm p}<\rvir)}\, .
\end{equation}
In this case SCCs are defined by $C_{\rm{scaled}} > 0.5$, WCCs fulfil $0.2 < C_{\rm{phys}} \leq 0.5$, while NCCs are defined by $C_{\rm{phys}}\leq 0.2$. \cite{santos2008} points out that introducing apertures that are scaled with $\rvir$ will naturally make lower redshift clusters appear more concentrated. 

To derive the X-ray emission of our clusters, we use the APEC collisional ionization models \citep{smith2001} applied to each gas cell, as a function of density, temperature, and metallicity, adopting solar abundances (following \textcolor{blue}{Nelson et al. submitted}).

We always take two-dimensional projections through a line-of-sight depth equal to $\pm r_{\rm 200c}$. The view direction is random with respect to the orientation of each cluster or BCG. Projections use the usual cubic-spline kernel approach. Star-forming gas, which is pressurized by our sub-grid ISM model, is assigned a temperature of 1000\,K appropriate for the mass-dominant cold phase, and so does not contribute to the X-ray signal. Throughout this paper, we consider the intrinsic X-ray luminosity of the clusters without accounting for observational effects. A visualization of the soft-band ($0.5 - 2$\,keV) X-ray maps of TNG-Cluster halos, and comparisons of X-ray scaling relations to observational data, are presented in the companion paper by \textcolor{blue}{Nelson et al. (submitted)}.


\section{The $z=0$ Cluster Population of TNG-Cluster} \label{sec_z0}

TNG-Cluster returns a sample of 352 galaxy clusters with a median mass of $\mvir = 4.3\times 10^{14} \msun$ at $z=0$. At the current epoch, the largest (smallest) cluster has a mass of $\mvir = 1.9 \times 10^{15} \msun$ ($\mvir = 1.0 \times 10^{14} \msun$), whereas the $\mtc$ spans a range of $2\times 10^{14} \msun < \mtc < 2.5 \times 10^{15} \msun$, with a median value of $\mtc = 6.4 \times 10^{14} \msun$. Their central galaxies have an average stellar mass of ${\rm M}_{\rm star} =3.1 \times 10^{12} \msun $, spanning a range of $6.0 \times 10^{11} \msun \leq {\rm M}_{\rm star} \leq 9.2 \times 10^{12} \msun$. These central galaxies host SMBHs with average mass of ${\rm M}_{\rm BH} =9.9 \times 10^{9} \msun $, reaching masses up to a few $10^{10} \msun $. At $z=1$, the median mass of the TNG-Cluster sample is $\mvir = 1.0\times 10^{14} \msun$ and ranges from $9\times 10^{13} \msun < \mvir < 6.5\times 10^{14} \msun$. More details on the demographics of the systems in TNG-Cluster, and their satellites, are given by \textcolor{blue}{Nelson et al. (submitted)} and \textcolor{blue}{Rohr et al. (submitted)}.

\subsection{Maps of thermodynamical Properties}\label{subsec_thermoProp}

We begin our exploration of TNG-Cluster by visualizing the gas density (Figure~\ref{fig:CoreGallery_ne}), entropy (Figure~\ref{fig:CoreGallery_K}), temperature (Figure~\ref{fig:CoreGallery_T}) and cooling time (Figure~\ref{fig:CoreGallery_tcool}) in central regions of the simulated galaxy clusters. We select 16 objects to illustrate the large diversity of cluster cores and the processes that shape their morphology and structure. We include examples of SCCs, WCCs and NCCs (as labeled) based on central cooling time as per the definitions of Sec.~\ref{ch:CCcritDef} and Table~\ref{tab:CCcuts}.

Each panel shows a thin slice of the central 0.2 $\rvir$, projected along a line of sight of 15\,kpc. We select the radius 0.2 $\rvir$, to focus on the central cores, roughly separating the regions influenced by non-gravitational processes from the regions where the thermodynamical profiles are self-similar \citep[see discussion below and e.g.][and references therein]{mcdonald2017}. For the systems shown, 0.2 $\rvir$ spans from 211\,kpc to 565\,kpc. Note that these maps deliberately exclude the gas associated to cluster galaxies that are not the central.

Figure~\ref{fig:CoreGallery_ne} visualizes the spatial distribution of the electron number density. For most clusters the density rises towards the core, reaching $n_{\rm e} \gtrsim 0.1 \rm{cm}^{-3}$ in the center.

However, the phenomenology can be diverse. For example, in the second panel of the top row, we see the merging of two clusters, with mass ratio of ${\rm M}_{\rm main}/{\rm M}_{\rm sub} = 154.25$. In the next panel to the right, a satellite 
is falling into the cluster, leaving a tail of stripped material and creating a bow shock, in analogy to jellyfish galaxies seen in TNG100, which are naturally captured in these simulations \citep{yun2019}. The population of massive satellites in TNG-Cluster and their circumgalactic media are studied in the companion paper by \textcolor{blue}{Rohr et al. (submitted)}. 

In many clusters, shell-like structures or rings are visible. These shells are likely due to AGN feedback~\citep{pillepich2021, fabian2006}, but can also arise due to sloshing \citep{zuhone2019, sanders2020}. Within the TNG-Cluster model, AGN feedback can create low density/high temperature bubbles close to the BCGs \citep{nelson2019b}. Such bubbles are over-pressurized and underdense, and can likely buoyantly rise through the ICM volume \citep{zhang2018}. On the other hand, features like swirls are evident at the boundary of different phases, possibly due to Kelvin-Helmholtz or Rayleigh-Taylor instabilities in the ICM \citep{gaspari2013a, roediger2013}.

Figure~\ref{fig:CoreGallery_K} shows the mass-weighted entropy distribution of the gas. We can clearly see a connection between the entropy in the cluster cores and the cluster cool-core state, as indicated by the labels in each panel: the low-entropy cores of the SCCs are often well visible. In contrast, WCCs and NCCs do not typically feature such clear or prominent cores in entropy. Edges in entropy trace features in the electron number density, with low-entropy regions corresponding to high-density regions. In SCC clusters stronger variations in entropy are manifest, while entropy gradients are flatter in NCC clusters. This can be related to the presence of turbulence in the ICM: strong turbulent diffusion leads to effective mixing of multi-phase gas in the ICM, which flattens the entropy gradients \citep{gaspari2013}.

In Figure~\ref{fig:CoreGallery_T} we show projections of gas mass-weighted temperature, for the same halos and on the same scale as in previous Figures. Unlike entropy, the temperature in the central region strongly depends on the mass of the halo. In the most massive halos, temperatures can reach values $>10^{8.3} $\, ($> 17.2$\,keV), while clusters at the low-mass end reach average temperatures in the core of $\sim 10^{7.4}$\, ($\sim 4.5$\,keV). The two clusters with the highest temperature are the most massive halo, and the merging halo. In addition to the overall dependence on cluster mass, core structures are visible in the morphology of the gas for the SCCs. 

A few of the depicted clusters feature cold, clearly rotating gaseous disks at their centers, with characteristic sizes of $\sim 40-50$\,kpc (lower left panel), while others have no such features. By visual inspection of the 16 presented TNG-Cluster systems at $z=0$, we find no obvious distinction between SCCs and NCCs regarding disk presence.

Lastly, Figure~\ref{fig:CoreGallery_tcool} shows maps of mass-weighted cooling time, whose spatial morphology closely resembles that seen in entropy. It is visually evident from Figure~\ref{fig:CoreGallery_tcool} that NCCs do not have low cooling time core regions. Instead, their cooling times reach $>10^{1.5}$\,Gyr, and have little variation in their central morphologies. On the other hand, clear core-like structures with low cooling times can be found in clusters classified as SCCs, whereby the central regions still exhibit a wider diversity. In their cores the cooling time reaches values $< 1$\,Gyr, in some cases the cooling time is even $\ll 0.1$\,Gyr. The variety in WCCs is even larger: some WCCs have cooling times of $< 100 $\,Myr in the central region, while others have cooling times $> 1$\,Gyr.

\begin{figure}
    \centering
    \includegraphics[width=0.48\textwidth]{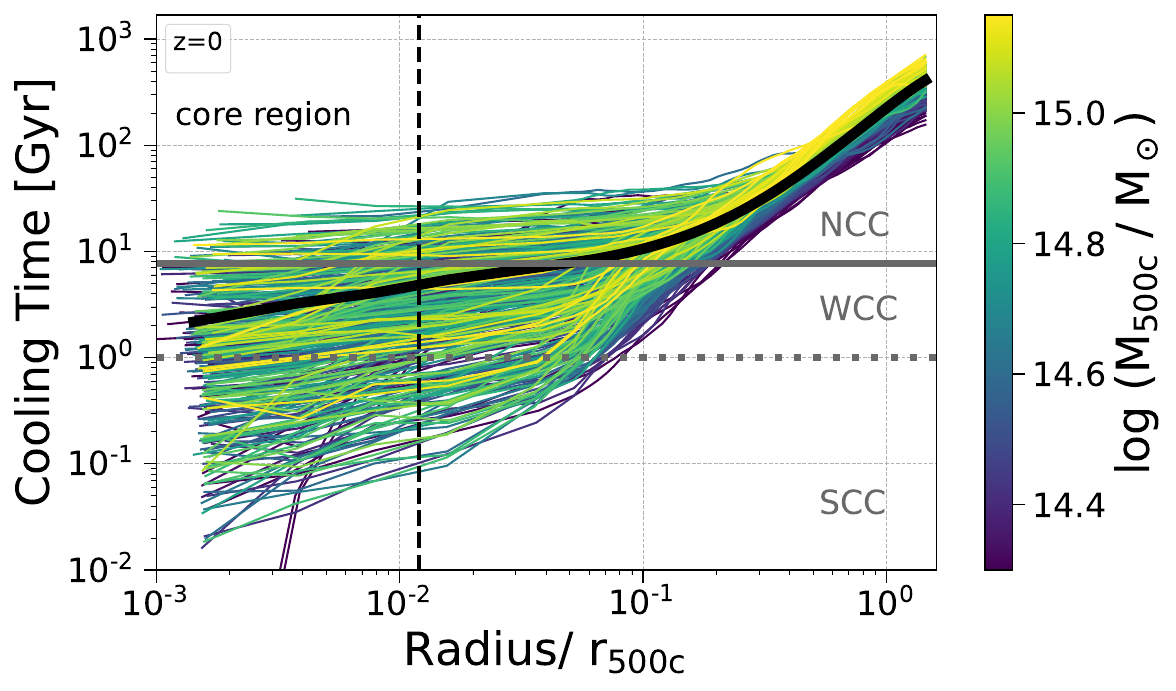}
    \includegraphics[width=0.48\textwidth]{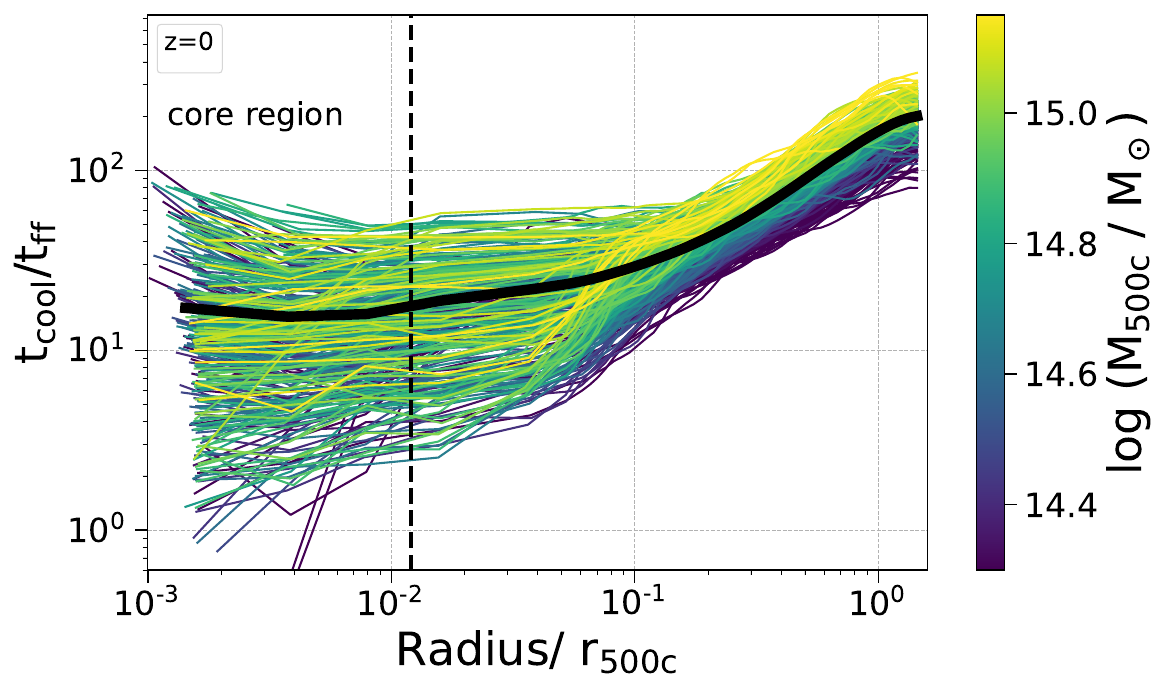}
    \caption{3D radial profiles of the cooling time (top) and ratio of cooling time to free-fall time (bottom) for all clusters in the TNG-Cluster sample at $z=0$, with focus on the central regions. The thin curves denote individual clusters and are color-coded by $\mvir$. The black curve shows the mean profile across all simulated clusters.}
    \label{fig:Profilestcoolz0}
\end{figure}

\begin{figure*}
    \centering
    \includegraphics[width=0.48\textwidth]{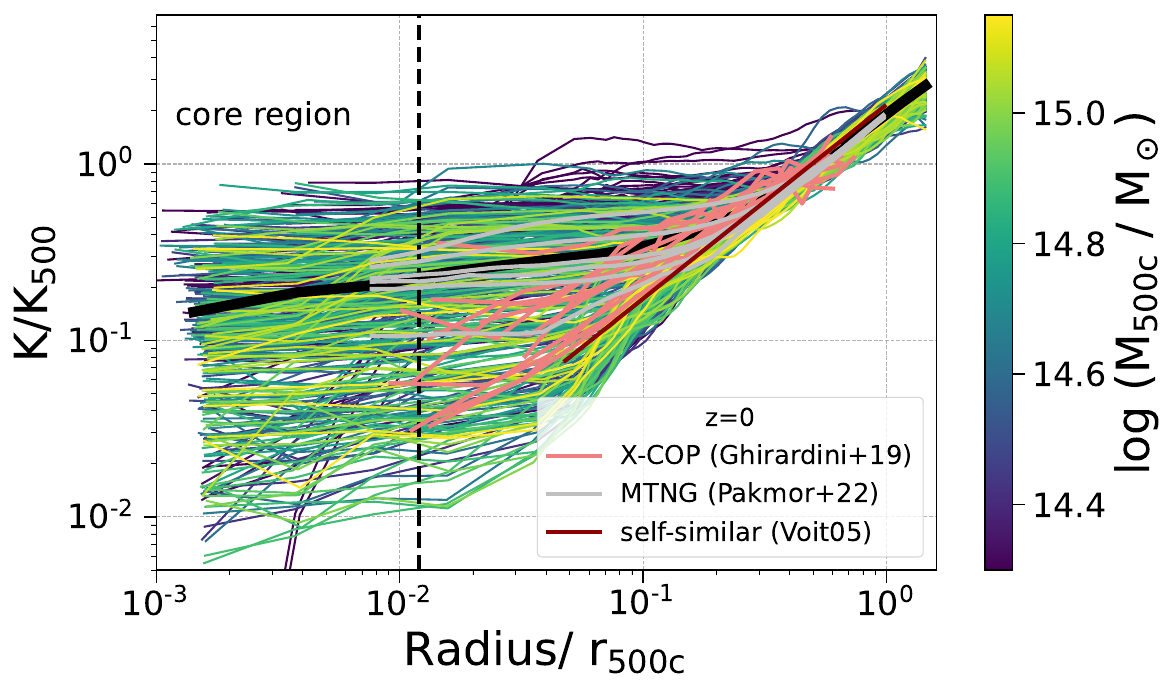}
    \includegraphics[width=0.48\textwidth]{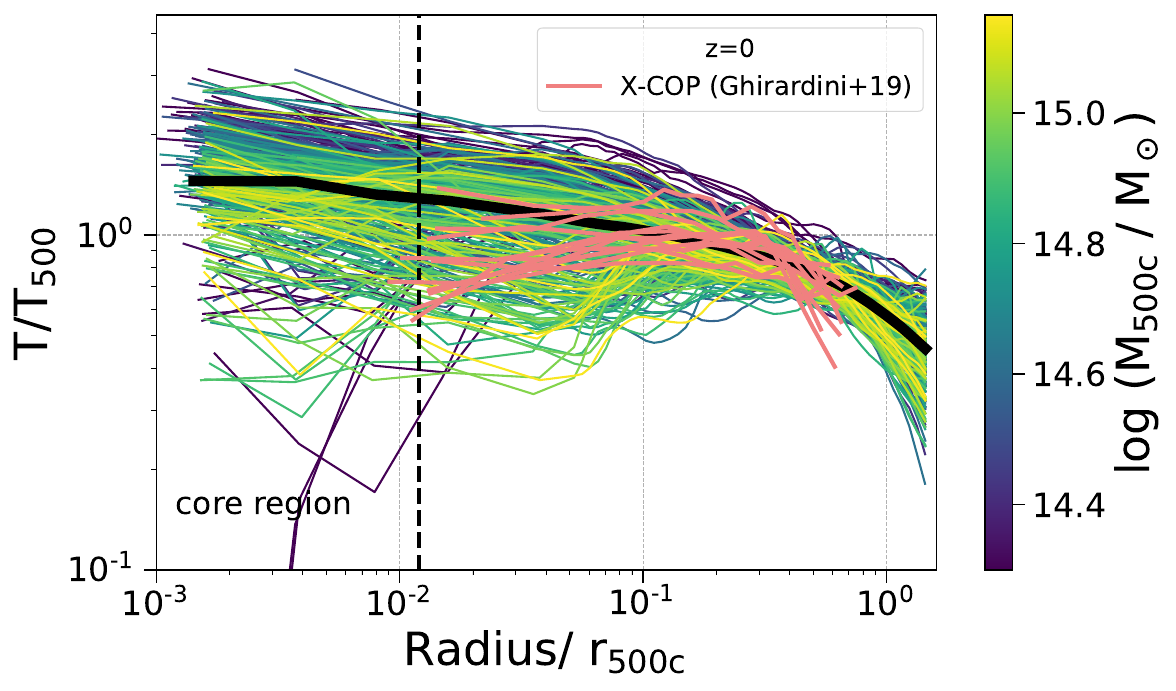}
    \includegraphics[width=0.48\textwidth]{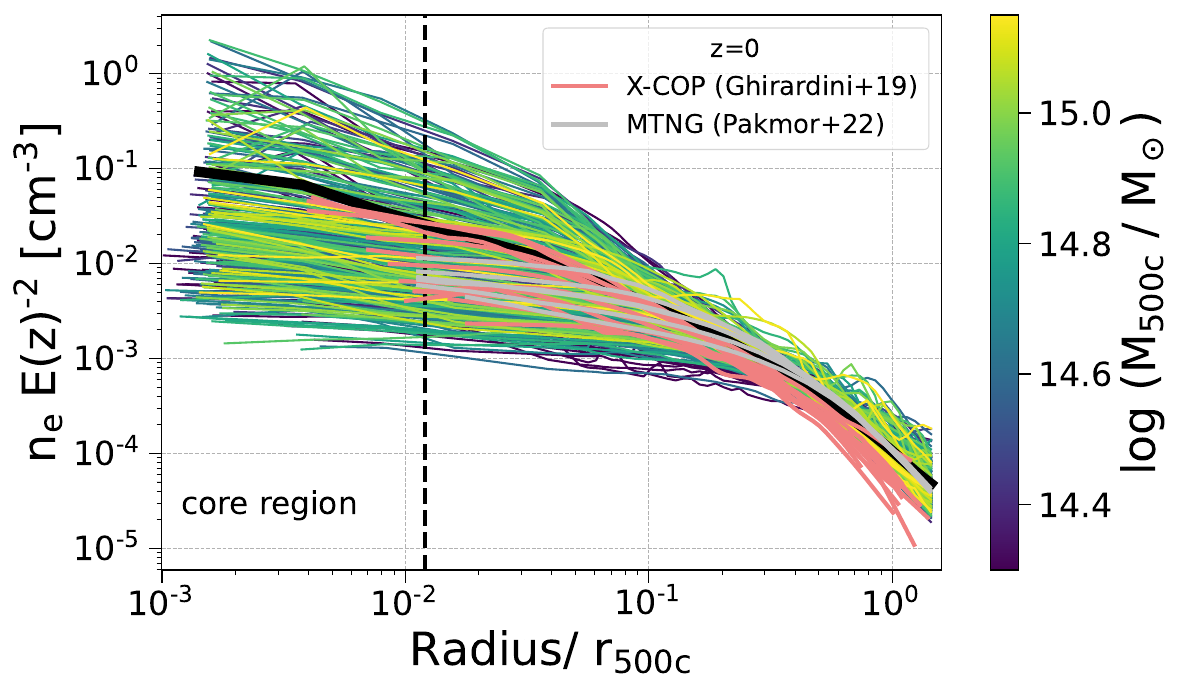}
    \includegraphics[width=0.48\textwidth]{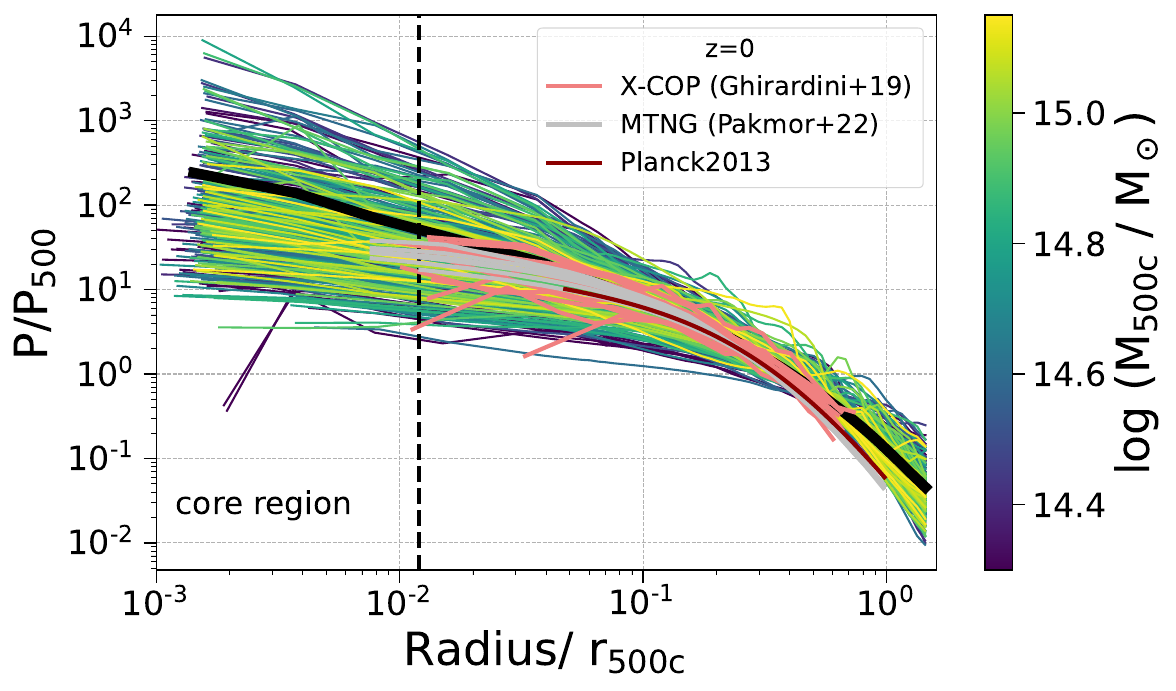}
    \caption{Individual and mean 3D radial profiles of all clusters in the TNG-Cluster sample at $z=0$. We show mass-weighted, radially-averaged entropy (upper left), temperature (upper right), electron number density (lower left), and pressure (lower right). We normalize by $\rvir$ and show log distance, to focus on the central regions. Where applicable, we normalize the profiles by the virial quantity. The black curves show the mean profile across the entire cluster population, whereas the individual curves are color-coded by M$_{\rm 500c}$. 
    We compare to the simulated profiles of MilleniumTNG \protect \citep[][grey lines]{pakmor2023} and the observed profiles of \protect \cite{ghirardini2019} (orange lines). The $\rvir$ values for profiles taken from the literature are rescaled by $\sim 0.68$ to account for different methodologies (see Sec \ref{subsec_CCvsNCC}). We see that the profiles are sorted by mass with low scatter at larger radii, suggesting that thermodynamical properties in the outer regions are set by global quantities, namely, mass. In contrast, the profiles have larger scatter in the center and are no longer ordered by mass, suggesting other processes strongly influence cluster evolution. TNG-Cluster produces realistic galaxy clusters in terms of their radial profiles, which at face value are in reasonable agreement with observations and other numerical simulations.}
    \label{fig:ProfilesNormedKTnePz0}
\end{figure*}

\subsection{Radial Profiles of Cooling Times}

Beyond the spatial complexity of Figure~\ref{fig:CoreGallery_tcool} and irrespective of CC status, the 3D spherically-averaged radial profiles of the ICM cooling time increase inside out: see Figure~\ref{fig:Profilestcoolz0}, top panel. 

In the central regions, the scatter among the individual clusters is large, spanning a range of $2\times 10^{-2} < t_{\rm cool}/\rm{Gyr} < 20$. In the outskirts, the scatter is smaller ($t_{\rm cool}(\rvir)$ = 150 - 700 Gyr) and set by the mass $\mvir$ of the halos: cooling times are larger in more massive clusters. Based on our fiducial definitions (Sec.~\ref{ch:CCcritDef} and Table~\ref{tab:CCcuts}), CCs are those with central cooling time shorter than one billion years. Typically, NCCs have flatter profiles in the core, whereas SCCs often exhibit a roughly power-law profile.

We expect that a multi-phase core can form if the ratio between cooling time and free-fall time is $t_{\rm cool} / t_{\rm ff} \lesssim 10$ \citep{sharma2012, voit2017}. In this case, cold dense gas condensates out of the hot ICM. The profiles of $t_{\rm cool} / t_{\rm ff}$ for all individual halos in TNG-Cluster have smaller ratios in the center than at $\rvir$: Figure~\ref{fig:Profilestcoolz0}, bottom panel. In the core the cluster-to-cluster variation is again larger ($t_{\rm cool} /t_{\rm ff}  =  1 - 100$), while it is smaller at $\rvir$ ($t_{\rm cool} /t_{\rm ff}  = 80-300$). It is evident from Figure~\ref{fig:Profilestcoolz0}, bottom panel, that a large fraction of systems in TNG-Cluster at $z=0$ satisfy the condensation condition ($t_{\rm cool} /t_{\rm ff} < 10$) also in the radial averages, in addition to more localized patches of cooling gas that can be evinced from the maps: hence, a non-negligible fractions of TNG-Cluster objects are actually CCs, as we quantify below.

\subsection{Radial Profiles of thermodynamical properties}\label{subsec_thermoProf}

Despite the morphological richness showcased in Figures~\ref{fig:CoreGallery_ne}-\ref{fig:CoreGallery_tcool}, the 3D spherically-averaged radial profiles of thermodynamical properties are well behaved and broadly consistent with previous models and available observational constraints. We show this in Figure~\ref{fig:ProfilesNormedKTnePz0} for all the 352 clusters of TNG-Cluster at $z=0$, by focusing on quantities that can be in principle inferred from observations:
entropy (upper left), temperature (upper right), density (lower left), and pressure (lower right). 

These quantities are normalized to the corresponding virial value, if applicable. We measure the profiles using 50 logarithmically spaced radial bins in the range of 10$^{-3}$ to 1.5 $\rvir$. Each curve is colored by $\mvir$ of the cluster, from low-mass (dark blue) to high-mass (yellow). The thick black line shows the mean profile of the entire TNG-Cluster population at $z=0$.

On average the entropy increases from the core to the outskirts. Some profiles are flat in the core -- these are typically the NCCs. Other profiles, typically the SCCs, have dips in the core region. We note that in several cases, including un-normalized entropy, radial profiles are roughly ordered by halo mass at $\rvir$, i.e. there is a clear systematic trend between entropy in cluster outskirts and mass, but this relationship breaks down in core.

The cluster-to-cluster variation of the profiles increases towards the center and reaches two orders of magnitude in the core. This implies that the outer parts of the profiles are set by gravitational processes, while the center is influenced by non-gravitational processes (AGNs, stellar feedback, cooling).  Only at $r = 0.6 \rvir$ the entropy profiles approach the self-similar expectations of \citep{voit2005}. A small fraction of all clusters (less than 1\%) show entropy as low as $\sim 10^{-2}$\,keV\,cm$^2$ in their cores, which could be due to e.g. the mis-centering of the halo, and/or to the disturbed nature of the systems.

We compare the TNG-Cluster profiles to those of MilleniumTNG \citep[MTNG;][]{pakmor2023}, for their 0.2\,dex mass bins in the range $14 < \log_{10}({\rm M}_{500}/\msun) < 15.2$ (gray curves). They agree well, as expected given that MTNG and TNG-Cluster are based on the same galaxy formation model, barring the absence of magnetic fields in the former. We also compare at face value to the profiles derived observationally in the X-COP project by \cite{ghirardini2019}, for twelve clusters in the $3.48 \times 10^{14} - 8.95 \times 10^{14} \msun$ mass range. 

The TNG-Cluster profiles are in reasonable agreement with the observed profiles, in the sense that they all overlap in parameter space. The X-COP profiles have average entropy values towards the central regions towards the lower side of TNG-Cluster values. This could be due to selection effects in the observed sample, to the fact that we compare intrinsic to X-ray derived entropy, or to modeling issues. Indeed, the lack of power-law i.e. cool-core entropy profiles has been a common issue of past simulations of galaxy clusters \citep{barnes2017a}, although it is not clear if this is due to low numerical resolution, missing, or e.g. overly simplified physical models \citep{altamura2023}.

The profiles of the other three quantities in Figure~\ref{fig:ProfilesNormedKTnePz0} (temperature, density and pressure) behave similar to the entropy profiles. 

At $\rvir$, the profiles scatter less than in the centers and are roughly ordered by mass. At $\sim 0.3 \rvir$, the halo-to-halo variation of all three physical quantities increases significantly, reaching a scatter of one order of magnitude (temperature) or even three orders of magnitude (density and pressure). The scatter in central density does not depend on halo mass (see also Figure~\ref{fig:CCcritVsM500}) and it induces the large scatter seen in entropy and pressure. 
Overall, the profiles in the central regions are ordered in terms of core status. We speculate that the small scatter in the temperature profiles results because heating at such high gas temperatures is rapidly offset by efficient cooling, or simply because the impact of AGN feedback on gas density versus temperature differs.

\begin{figure*}
    \centering
	\includegraphics[width=\textwidth]{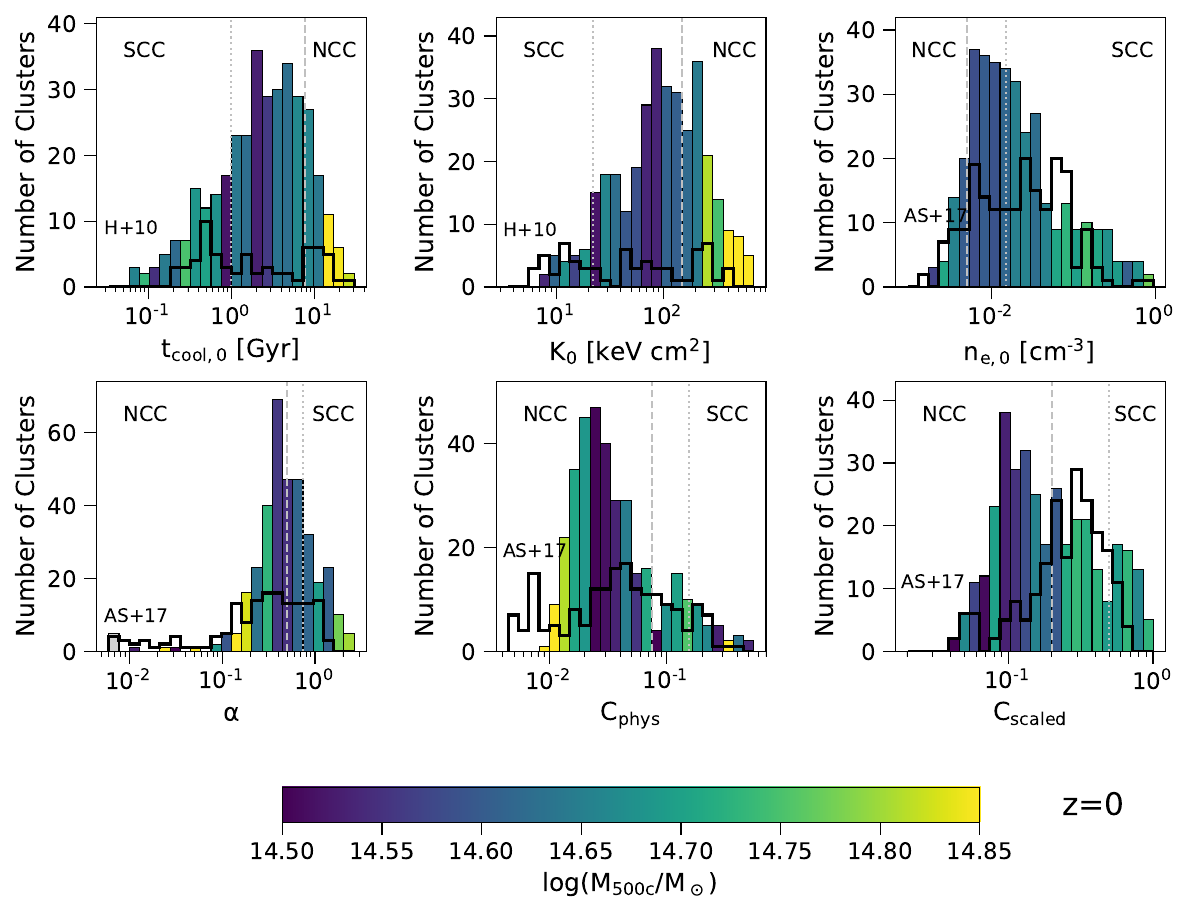}
    \caption{Distribution of the six CC criteria that we consider for all 352 halos of TNG-Cluster at $z=0$. From upper left to lower right, these are: central cooling time, central entropy, central electron number density, cuspiness parameter, physical X-ray concentration, scaled X-ray concentration. In each case, the SCC (NCC) thresholds are indicated by the dotted (dashed) vertical lines. Each of the 20 bins is colored by the mean $\mvir$ of all clusters in the bin. The left-most (grey) bin in the histogram of the cuspiness parameter indicates all cluster that have $\alpha < 0$. The histograms indicated by the black lines are the results from \protect\cite{andrade-santos2017, hudson2010} (see text for details). The comparison is meant to be only qualitatively, as we do not replicate the measurement procedure of the observational data, In all six cases, TNG-Cluster produces continuous distributions of CC criteria with no clear bimodalities. The thresholds from previous work and observational findings do not clearly coincide with any features in our distributions. }
    \label{fig:HistCCcrit}
\end{figure*}

The electron number density and the pressure profiles agree well with the results of MTNG \citep{pakmor2023}. For the pressure profiles we also plot the results of the \textit{Planck} ESZ sample \citep{planckcollaboration2013}, with a mass range of $9\times 10^{13} - 1.5\times 10^{15}\msun$. Their average pressure profile is consistent with the average of TNG-Cluster. This lends credibility to the outcome of TNG-Cluster even though this comparison is also at face value, i.e. without replicating selection effects nor observational measurements. The observed profiles of the X-COP sample \citep{ghirardini2019} also agree well with the simulated profiles at $r >0.1 \rvir$. At smaller radii, on the other hand, the observed temperature profiles tend to have a different shape than the simulated profiles, dropping to smaller temperatures. On average, the X-COP observed profiles have lower densities in the central regions. This may occur because the X-COP sample, like other \textit{Planck}-selected samples, is dominated by NCCs \citep{ghirardini2019}. The X-COP observed pressure profiles agree well with those predicted by TNG-Cluster.

Overall, Figure~\ref{fig:ProfilesNormedKTnePz0} demonstrates that TNG-Cluster produces reasonably realistic galaxy clusters, in that their radial profiles are broadly consistent with those from observations and other numerical simulations. We therefore proceed to study the CC and NCC populations produced by TNG-Cluster.

\subsection{Cool-Cores vs. Non-Cool-Cores at $z=0$}\label{subsec_CCvsNCC}

Figure~\ref{fig:HistCCcrit} shows the $z=0$ distributions of central thermodynamical properties for all 352 clusters for the six CC criteria introduced in Sec.~\ref{ch:CCcritDef}. The thresholds for SCCs (NCCs), stated in Table~\ref{tab:CCcuts} are indicated by the dotted (dashed) vertical lines. We also color each bin by the mean halo mass of clusters in that bin.

TNG-Cluster produces continuous distributions of the properties for all six CC criteria. These span $\sim 1-2$ dex across the entire TNG-Cluster sample, but are highly asymmetric. Except for a hint of a  bimodality in the distributions of the physical and scaled concentration parameter, no clear bimodality is visible in the distributions. We have checked and this remains the case also when e.g. considering only clusters in smaller mass bins. The only exception is a hint of bimodality in central cooling time, central entropy, scaled and physical concentration parameter for the Perseus mass range ($\log_{10} (\mvir/\msun) = 14.40 - 14.87$), although it is not clear if they are statistically significant.

The histograms of Figure~\ref{fig:HistCCcrit} are unimodal and indicate that, at least according to TNG-Cluster, the SCCs and NCCs are the extremes of each distribution. The figure also clearly shows that the thresholds used to classify core states, which are observationally motivated, do not reflect any particular feature in the simulated distributions and hence appear somewhat arbitrary. This is consistent with similar claims based on TNG300 and lower-mass clusters \citep{barnes2018}. In fact, the main physical difference between TNG-Cluster and TNG300 is that the two have different cluster mass distributions: the high-mass (full) sample of \cite{barnes2018} has a median mass of ${\rm M}_{500} = 2.7 \times 10^{14} \msun$ (${\rm M}_{500} = 8.8 \times 10^{13} \msun$), while our full TNG-Cluster sample is 1.6 times more massive. The distributions of all six central ICM properties of TNG300 are also continuous and unimodal \citep{barnes2018}, with similar shapes as in TNG-Cluster, which confirms those findings with $\sim 10$x better statistics.\footnote{In Appendix.~\ref{sec_appendix_newCCcrit}, we quantify percentiles of these continuous distributions and suggest their use as a new approach for the classification of cluster core states.}

According to TNG-Cluster and considering the whole cluster sample in the $1\times 10^{14} - 2 \times 10^{15} \msun$ mass range, the mean values of cooling times, central entropy, central electron number density, cuspiness, physical concentration parameter, and scaled concentration parameter are $\bar{t}_{\rm{cool}, 0} = 4.16$\,Gyr, $\bar{K}_0 = 137.4$\,keV cm$^2$, $\bar{n}_{\rm{e}, 0} = 5.6\cdot 10^{-2}$\,cm$^{-3}$,  $\bar{\alpha} = 0.6$, $\bar{C}_{\rm{phys}} = 0.057$, and $\bar{C}_{\rm{scaled}} = 0.25$, respectively. However, barring the scaled concentration parameter, the distributions of all central ICM physical properties are highly asymmetric, with long tails towards systems with short cooling times, low entropy values, high central densities, flat and high-concentration cores. Moreover, clusters with higher central cooling times have slightly larger masses than clusters with lower central cooling times; higher central entropy clusters are more massive. Systems with {\it flatter} and more concentrated centers (not rescaled) are also at the high-mass end of the TNG-Cluster population, although there is no strong monotonic mass trend across the whole parameter space. Finally, the distribution of central electron number density shows no clear mass trend.

With more limited statistics and a smaller mass range, \cite{barnes2018} could identify only weak or no mass trends for the central ICM properties of TNG300 clusters. In comparison, the values of TNG-Cluster for central cooling time, entropy, and density confirm the mass trends suggested in \cite{barnes2018}, but with mean values shifted more towards the SCC regime. The mean cuspiness parameter of TNG-Cluster also shifts towards the SCC regime, whereas the mean scaled concentration parameter is in agreement with the full sample of \cite{barnes2018}, although smaller than the value of their high-mass sample. Beyond differences in the underlying mass distribution of the samples, discrepancies in the resulting distributions of central ICM properties can be due to differences in the measurements. For example, even in simulated data, the X-ray concentrations may differ due to differences in X-ray emission modeling, or in the extraction of the concentration parameters from the X-ray maps.

Keeping in mind the systematic uncertainties mentioned above, which can bias the comparison among analyses, we discuss now how the TNG-Cluster distributions of Figure~\ref{fig:HistCCcrit} compare to those derived observationally, when available. As we do not replicate the measurement procedures to extract ICM central properties as in observations, and given the diverse cluster samples, our qualitative comparison is intended at face value.

Observed distributions of central cooling time and central entropy are available for the sample of 64 X-ray flux-limited clusters by \cite{hudson2010}. Their sample spans a mass range of $2.95\times 10^{13} - 3.63 \times 10^{15} \msun$. The mean central cooling time (entropy) of their sample is with $\bar{t}_{\rm cool} = 4.86$\,Gyr ($\bar{K}_0 = 99.2$\,keV\,cm$^2$) comparable to (lower than) ours. Interestingly, according to \cite{hudson2010}, both histograms appears trimodal instead of unimodal as in the simulations, with TNG-Cluster spanning a similar range of $t_{\rm{cool},0}$ and $K_0$ values. 

We can qualitatively compare the distributions of central number density, slope and concentration parameters to the results of the \textit{Planck} Early Sunyaev-Zel’dovich (ESZ) survey from \cite{andrade-santos2017} containing 164 clusters with $z\leq 0.35$: black solid histograms in Figure~\ref{fig:HistCCcrit}. Their average mass is slightly larger than ours, but we have a similar number of high-mass clusters. The distribution of central densities inferred in the ESZ survey has a mean value of $\bar{n}_{\rm{e}, 0} =1.07 \times 10^{-2}$\,cm$^{-3}$ and does not show a clear dichotomy: it spans a similar range of central densities compared to TNG-Cluster but it is more symmetric. The mean value of the cuspiness of ESZ clusters is somewhat smaller than in TNG-Cluster ($\bar{\alpha} = 0.4$ vs. $0.6$), resulting in a larger NCC fraction in the observed sample. The observed and simulated distributions both have a tail towards smaller values, but the tail is more prominent in the observed distribution, causing the mean value to be smaller. The profiles used to compute the cuspiness parameter are measured from \textit{Chandra} X-ray data: even though we use the same definition of the cuspiness parameter as \cite{andrade-santos2017}, we do not account for observational realism effects in the X-ray surface brightness maps.

\begin{figure*}
    \centering
	\includegraphics[width=0.46\textwidth]{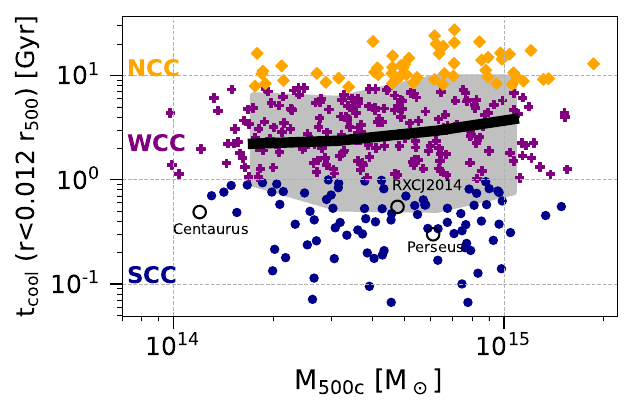}
	\includegraphics[width=0.46\textwidth]{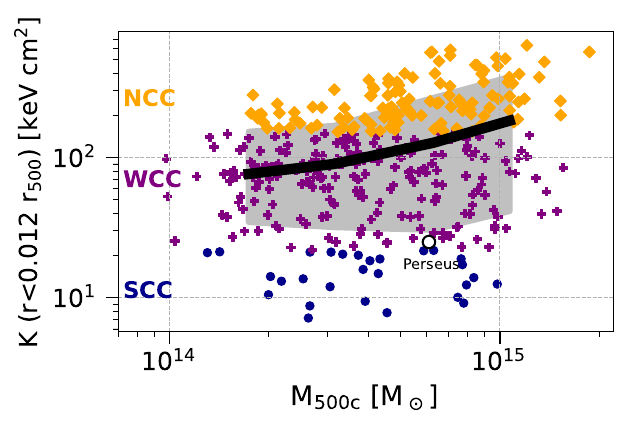}
	\includegraphics[width=0.46\textwidth]{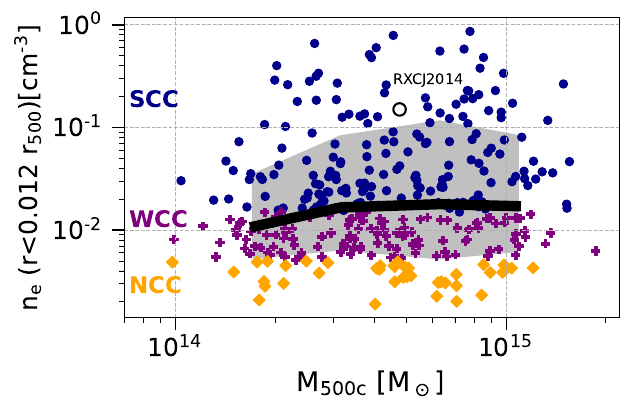}
	\includegraphics[width=0.46\textwidth]{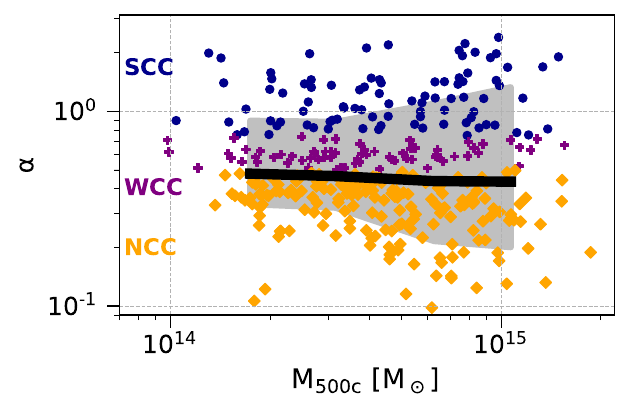}
    \includegraphics[width=0.46\textwidth]{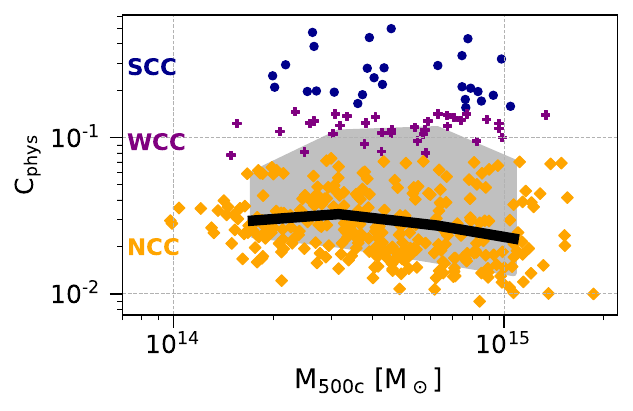}
	\includegraphics[width=0.46\textwidth]{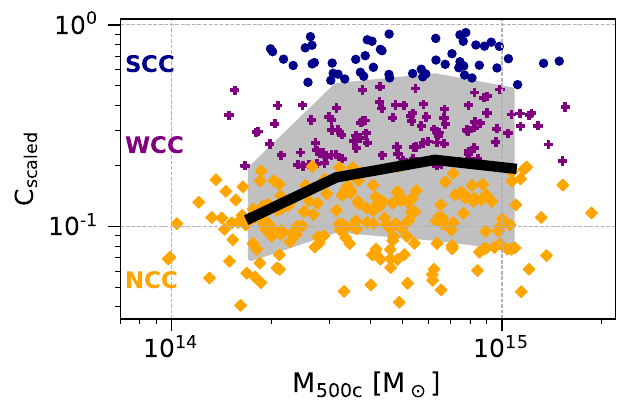}
    \caption{CC criteria as a function of $\mvir$ for the entire TNG-Cluster sample at $z=0$. The color of the symbols indicates whether the clusters are classified as SCC (dark blue dots), WCC (purple crosses) or NCC (orange diamonds) according to the quantity on the y-axis. The black line shows the median trend of the CC criteria with $\mvir$. The silver solid region shows the 16 to 84 percentile ranges. From upper left to lower right: central cooling time, central electron number density, central entropy, physical concentration parameter, scaled concentration parameter, cuspiness parameter. We also make qualitative, at face value only, comparisons to values from selected, nearby clusters (see text). Overall, there is no well-defined mass trend visible across all CC criteria. In some cases, central cluster properties do not depend on mass, while other properties show increasing, decreasing, or non-monotonic behavior with halo mass.
    The number of SCC cluster varies substantially depending on the defining criteria.}
    \label{fig:CCcritVsM500}
\end{figure*}

Finally, whereas the mean value of the physical concentration parameter of ESZ clusters is close to that of TNG-Cluster ($\bar{C}_{\rm{phys}} = 0.043$ vs. $0.057$), the observed systems exhibit significantly larger values of the scaled concentration ($\bar{C}_{\rm{scaled}} = 0.51$ vs. $0.25$). This occurs because the TNG-Cluster distribution has a peak at lower values, while the data peak higher. Since the average simulated $C_{\rm phys}$ match well with the observed one, the difference in $C_{\rm scaled}$ is likely due to differing radii within which the concentration is computed. Higher concentrations are obtained in smaller apertures, and \cite{andrade-santos2017} compute $r_{\rm 500}$ from $M_{\rm 500}$, whereas we use the $\rvir$ spherical-overdensity value. These are by a factor of $\sim 1.47$ larger than the $\rvir$ values computed from $\mvir$. Thus, the smaller mean value of the simulated $C_{\rm scaled}$ may be due to this methodological mismatch.

From these comparisons, we conclude that simulated and observed values of the ICM core properties occupy similar parameter spaces, lending credibility to the CC/NCC fractions predicted by TNG-Cluster. Before presenting these fractions, we first finish our overview of TNG-Cluster core properties by further quantifying their cluster mass dependencies.

\subsection{Mass Trends of Core Properties}\label{subsec_CCcritVsM500}

\begin{figure*}
    \centering
    \tabskip=0pt
    \valign{#\cr
      \hbox{\begin{subfigure}[b]{.66\textwidth}
        \centering
        \includegraphics[width=\textwidth]{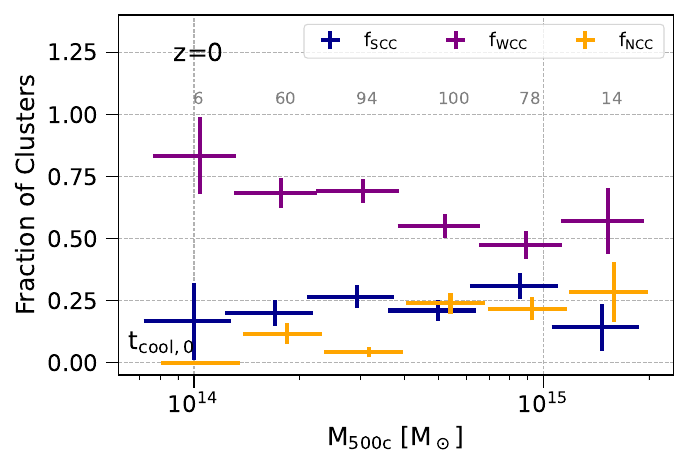}
        \end{subfigure}}\cr
      \noalign{\hfill}
      \hbox{\begin{subfigure}{.33\textwidth}
        \centering
        \includegraphics[width=\textwidth]{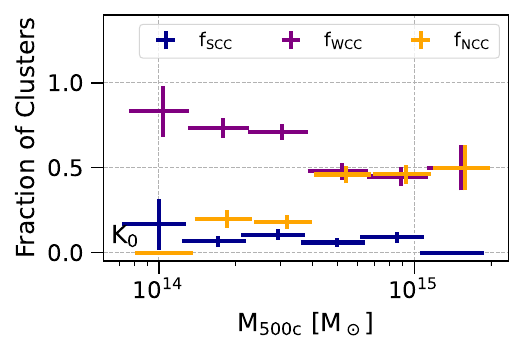}
        \end{subfigure}}
        \vfill
      \hbox{\begin{subfigure}{.33\textwidth}
        \centering
        \includegraphics[width=\textwidth]{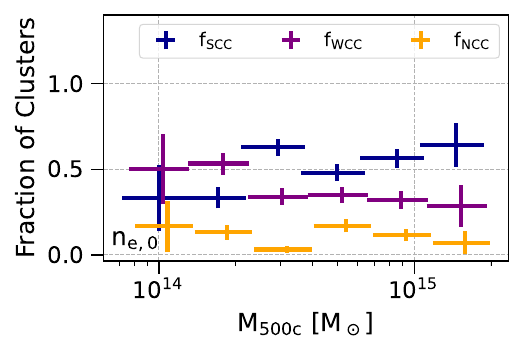}
        \end{subfigure}}\cr}
    \hbox{\begin{subfigure}{.33\textwidth}
        \centering
        \includegraphics[width=\textwidth]{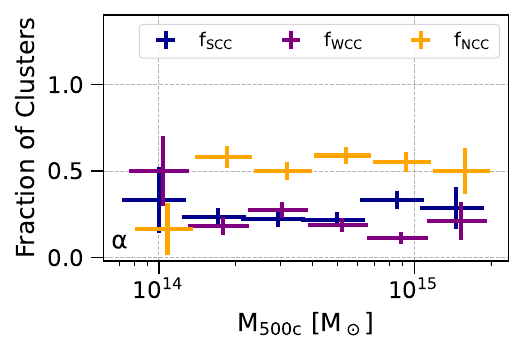}
        \end{subfigure}}
    \hbox{\begin{subfigure}{.33\textwidth}
        \centering
        \includegraphics[width=\textwidth]{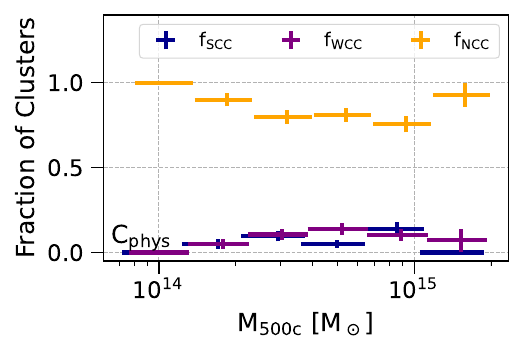}
        \end{subfigure}}
    \hbox{\begin{subfigure}{.33\textwidth}
        \centering
        \includegraphics[width=\textwidth]{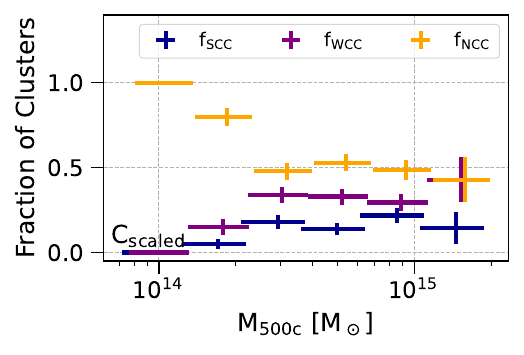}
        \end{subfigure}}%
    \caption{Fraction of SCC (dark blue), WCC (purple) and NCC (orange) clusters as a function of M$_{\rm 500c}$ in the TNG-Cluster simulation at $z=0$. We compute these fractions in 6 mass bins (bin size $\sim 0.2 $ dex). The uncertainties are computed by bootstrap re-sampling (10,000 times). We consider the same fiducial six CC criteria: $t_{\rm cool, 0}$ (main panel), $K_0$ (upper right), $n_{\rm e, 0}$ (center right), $\alpha$ (lower left), $C_{\rm phys}$ (lower center), and $C_{\rm scaled}$ (lower right).The respective CC criterion is indicated in the lower left corner of each panel. In the main panel, the grey numbers state the number of clusters in each bin. The cluster fractions vary considerably depending on the CC criterion, show varying trends with cluster mass and varying normalizations in cluster fraction.}
    \label{fig:fCCVsM500}
\end{figure*}

Figure~\ref{fig:CCcritVsM500} shows the same six quantities of the CC criteria as a function of cluster mass. We show the full TNG-Cluster sample at $z=0$: SCC clusters are represented by blue dots, WCCs as purple crosses, and NCCS as orange diamonds. The black solid curves show the medians, while the shaded gray areas give the 16 to 84 percentile ranges.

As already suggested by Figure~\ref{fig:HistCCcrit}, there are weak to no mass trends in core properties. For the central cooling time, the central entropy, and the scaled concentration parameter, the median increases slightly for larger cluster masses. The physical concentration parameter shows the opposite trend. In contrast, central electron number density and density exhibit relatively flat mass trends. At the same time, it is clear that the abundance of SCCs strongly depends on the defining criterion \citep[in agreement with][]{mcdonald2013, barnes2018}.

We add several data points to qualitatively demonstrate that TNG-Cluster produces core properties that are compatible with those of well-known observed clusters. The mass of Perseus is taken from \protect\cite{giacintucci2019a}, the central cooling time from \protect\cite{sanders2007} and the central entropy from \protect\cite{churazov2003}. The mass of Centaurus is taken from \protect\cite{walker2013} and its central cooling time from \protect\cite{edge1992}. The data for the strongest cool-core in the REXCESS sample, RXC J2014.8-2430, are from \protect\cite{haarsma2010a}. These real clusters lie in regions of the parameter space that are filled by numerous examples from TNG-Cluster, enabling us to study analogs of specific observed clusters. For example, in the companion paper by \textcolor{blue}{Truong et al. (submitted)}, we identify 30 Perseus-like analogs in TNG-Cluster and study their core kinematics in comparison to HITOMI and XRISM measurements.

\subsection{Cluster cool-core fractions}\label{subsec_fcc}

To distill the mass dependence of CC versus NCC clusters, we study a fundamental property of the cluster population: the (non) cool-core fractions. Figure~\ref{fig:fCCVsM500} shows the dependence of (N)CC fractions for TNG-Cluster as a function of halo mass at $z=0$. The SCC fraction is shown in blue, while the NCC (WCC) fraction is in orange (purple). We compute the fractions in six mass bins of $\sim 0.2$ dex each. The vertical bars give the uncertainty on the cluster fraction computed by bootstrapping, while the horizontal bars represent the width of each mass bin.

The main panel shows the cluster fractions using the central cooling time for classification. The SCC fraction (blue) shows a mild increase towards higher masses, from $\sim 20$\% to $\sim 30$\% across our mass range.\footnote{In the last bin the SCC fraction drops by 50\%, but due to low statistics, we treat the result in the first and last bin with caution. The first bin contains only 6, and the last one only 14 clusters.} In contrast, the WCC fraction rapidly decreases with mass, from 68\% to 47\% (excluding the outer bins). The NCC fraction generally increases with mass, from $\sim 10\%$ to $\sim 20\%$. The small panel on the top right shows the same SCC, WCC, and NCC cluster fractions, instead using central entropy as the defining criterion. We find the same trend as in the main panel, except that the SCC fraction remains low for all masses. The trends are, however, much more pronounced, i.e. the NCC fraction increases from $\sim 0\%$ to $\sim 50 \%$.

The trends in cool-core fractions when the classification is based on the central electron number density (right middle panel) are qualitatively different. The SCC fraction increases from 33\% to 56\%, while the WCC fraction decreases towards higher masses, and the NCC fraction fluctuates around $\sim10\%$. Physically, while central cooling time and entropy are closely related, it is clear that central density is modulated by other or additional processes, such that (N)CC fractions based on $n_{\rm e, 0}$ have qualitatively different mass trends. Different yet still, there is no mass trend in the cuspiness parameter $\alpha$ (lower left panel). The measured fractions are consistent with roughly constant values of SCC $\sim 26\%$, WCC $\sim 20\%$, and NCC fraction $\sim 54\%$.

Using the physical concentration parameter (lower center) to classify clusters, the number of NCCs decreases as a function of $\mvir$. While the overall decrease is large ($\sim 25\%$), the number of SCCs and WCCs increases only slightly with mass. The cluster cool-core fractions based on the scaled concentration parameter behave similarly, although the drop of the NCC fraction with mass is even larger ($\sim 50\%$). Although, both concentration parameters are computed from the same data, the differences in the definition lead to different mass trends and different CC fractions. As the aperture for the scaled concentration parameter increases with $\rvir$ and so $\mvir$,  this leads to different mass trends of the two concentration parameters.

Summarizing Figure~\ref{fig:fCCVsM500}, we find that the fraction of SCCs tends to increase with increasing halo mass. However, the fractions at fixed mass and the absolute change in fractions with mass are quite different for different CC criteria. These are sensitive to disparate properties of clusters and thus are sensitive to the various mass trends of the cluster properties, causing different mass trends of CC fractions. 

In the numerical work by \cite{burns2007} and \cite{planelles2009a}, a decreasing cool-core fraction with increasing mass is reported. However, a proper comparison is rather difficult as they use a CC criterion based on the central temperature drop and a different physics model. The simulation of \cite{planelles2009a} do not include metal dependent cooling and feedback mechanisms relevant for cluster physics. 

{\renewcommand{\arraystretch}{1.3}
\begin{table}[t!]
        \caption{Summary of the strong cool-core (SCC), weak cool-core (WCC), and non-cool-core (NCC) cluster fractions for the entire TNG-Cluster sample, combining all halo masses ($1.0 \times 10^{14} \msun < \mvir < 1.9 \times 10^{15} \msun$), at $z=0$. Each column shows one of the six different central criteria considered.}
         \begin{tabular}{lcccccc}
            \hline\hline
             & $t_{\rm{cool},0}$ & $n_{\rm{e},0}$& $K_0$&$C_{\rm{phys}}$ & $C_{\rm{scaled}}$&$\alpha$ \\
             \hline
             $f_{\rm{SCC}}$\hspace{-0.5em} & $24\pm 2$ &$52\pm3$ &$8\pm 1$ & $8\pm 1$ & $15\pm 2$ & $25\pm 2$ \\
             $f_{\rm{WCC}}$\hspace{-0.5em} & $60\pm 3$ &$37\pm3$ &$59\pm 3$ & $10\pm 2$ & $29\pm 2$ & $20\pm 2$ \\
             $f_{\rm{NCC}}$\hspace{-0.5em} & $16\pm 2$ &$11\pm2$ &$34\pm 3$ & $82\pm 2$ & $56\pm 3$ & $54\pm 3$ \\
             \hline\hline
        \end{tabular}
    \label{tab:fCC_wholeSample}
\end{table}}

We compare our findings to the mass trends of (N)CC fractions reported in \cite{barnes2018}. Using our fiducial CC criterion ($t_{\rm cool, 0}$), we find a increasing SCC and NCC fraction with increasing mass, and a decreasing WCC fraction. \cite{barnes2018} find the SCC fraction decreases, the WCC fraction stays constant, and the NCC fraction rises with mass. CC fractions based on $K_0$ are compatible, and show similar trends with mass. The mass trends of the cluster fractions using $n_{\rm e, 0}$, $\alpha$, $C_{\rm pyhs}$, and $C_{\rm scaled}$ of this work and \cite{barnes2018} are in reasonable agreement. TNG300 and TNG-Cluster employ the same galaxy formation model, although our sample has a larger median mass. Any differences in mass trends are therefore likely due to small details in the cluster classification procedure, numerical methodology in the CC criteria calculation, and the differing cluster masses of the samples.

Table~\ref{tab:fCC_wholeSample} summarizes our results for cluster (non) cool-core fractions of the complete TNG-Cluster sample at $z=0$. Overall, we find a SCC fraction of $24\pm 2\%$ ($t_{\rm cool, 0}$), $8\pm 1\%$ ($K_0$), $52\pm 3\%$ ($n_{\rm e, 0}$), $25\pm 2\%$ ($\alpha$), $8\pm 1 \%$ ($C_{\rm phys}$) and $15\pm 2\%$ ($C_{\rm scaled}$).

\begin{figure}
    \includegraphics[width=0.48\textwidth]{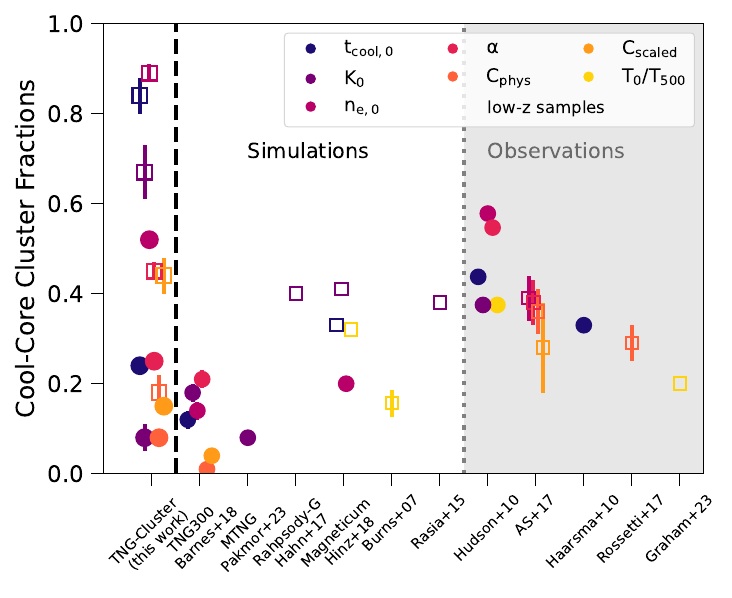}
    \caption{Cool-core fractions of clusters in TNG-Cluster (left), other simulations (center), and observations (right). Each are based on a given central thermodynamical property, as shown in the legend and color. Circles indicate `strong' cool-cores (SCC), while squares show cool-cores (CC). Note that we define the number of CC as SCCs+WCCs for this work. Even when using the same central property, the various works differ in many other aspects: the exact definitions used to define CC state, sample i.e. mean cluster mass and mass distribution, as well as the cluster redshift distribution. As a result, the comparison among simulations, and between simulations and the various observations, requires careful evaluation (see text) and are meant to be qualitatively, at face value only, in this figure.}
    \label{fig:fCCvsPaper}
\end{figure}

Figure~\ref{fig:fCCvsPaper} compares to other inferences from both observations (right side) as well as simulations (left side). The y-axis shows both strong cool-core fractions (circles) as well as cool-core fractions (squares), when available. Determinations from different CC criteria are given different colors, as labeled in the legend. To define the cool-core fraction we add the strong and weak cool-core fractions. However, this may result in larger CC fractions compared to other work, as various values between our thresholds for SCCs and WCCs are used to define cool cores in the literature. 

Broadly, observations have often inferred somewhat higher cool-core fractions than cosmological simulations of galaxy clusters. However, the exact definitions used to measure CC fraction, as well as sample properties including the cluster mass and redshift distributions and observational selection functions play a role. For this reason, the comparisons with other works are meant to be qualitative, at face value only. We discuss and compare to CC fractions from each of the six CC criteria below.

First, we compare to the predecessor of TNG-Cluster, the TNG300 simulation. \cite{barnes2018} compute the CC fraction for a sample of 370 clusters with ${\rm M}_{500}> 10^{13.75} \msun$, as well as separately for only a high-mass sub-sample of 49 clusters (${\rm M}_{500}> 2 \times 10^{14} \msun$). This high-mass sub-sample was created because the median mass of the complete simulated sample was significant lower than the masses from observed samples. Their SCC fractions using central cooling time and central electron number density are smaller than ours, for both their complete ($12\pm 2 \%$ [$t_{\rm cool,0}$] and $14\pm2 \%$ [$n_{\rm e,0}$]) and high-mass ($6\pm4\%$ [$t_{\rm cool,0}$] and $35\pm 6 \%$ [$n_{\rm e,0}$]) sample.

\cite{barnes2018} also use entropy to classify clusters, however they adopt the central entropy excess by fitting a power-law to the entropy profile. Despite this difference, the SCC fraction of their high-mass sample agrees with our SCC fraction within the uncertainties ($10\pm 6\%$ vs $8\pm 1\%$). Their SCC fraction based on the cuspiness agrees with ours ($21\pm 2\%$ vs $25\pm 2\%$) in the complete sample, and is lower ($10\pm 6\%$) in the high-mass sample. Similarly, the physical concentration parameter for their high-mass sample has a SCC fraction comparable to ours ($8\pm 6 \%$ vs $8\pm 1 \%$). However, our SCC fraction based on the scaled concentration parameter is half that of their high-mass sample.

Overall, the SCC fractions of TNG-Cluster are in good agreement with those from TNG300. For some CC criteria our sample leads to a larger SCC fraction, likely reflecting a (strong) halo mass dependence. 

The other key comparison is with the recent MilleniumTNG (MTNG) simulation \citep{pakmor2023}. This project also employs the TNG galaxy formation model, but with several minor modifications, including the exclusion of magnetic fields, and at somewhat lower numerical resolution than TNG300 or TNG-Cluster. At $z=0$ the fiducial MTNG run contains 9 galaxy clusters above $\mvir > 10^{15} \msun$ and more than 2000 clusters above $\mvir > 10^{14} \msun$. MTNG finds a SCC fraction of 8\% using the central entropy excess for ${\rm M}_{500}>10^{14}\msun$. However, they have zero SCCs in their high-mass sample of ${\rm M}_{500}>10^{15}\msun$ \citep{pakmor2023}. Applying, their entropy threshold we also find no cool-cores in such high-mass halos. However, there are CC metrics that lead to non-zero cool-core fractions at ${\rm M}_{500}>10^{15}\msun$ in TNG-Cluster. Interestingly, we clearly find that $K_0$ increases with halo mass \citep[as in TNG300;][]{barnes2019}, unlike in MTNG where \citet{pakmor2023} emphasize that central entropy decreases with increasing halo mass.

Using the Rhapsody-G simulation, \cite{hahn2017a} study 10 clusters with a mean mass of $\mvir = 5.8 \times 10^{14}\,\msun$. That work classifies a cluster as CC when the mean central entropy within 10kpc is less then 40 keV cm$^2$. They find a CC fraction of 44\%. If we employ the same entropy threshold, we find a CC fraction of 20\%. Although this is notably smaller, we adopt an aperture of 0.012 $\rvir$, while they use a fixed aperture of 10\,kpc, roughly twice as small as ours, leading to systematically higher central entropy in our case. This emphasizes that quantitative cross-simulation comparisons are difficult and will benefit in the future from unified analyses on e.g. publicly available simulations.

\cite{hinz2018} compute (S)CC fractions using central electron number density ($f_{\rm SCC} = 20\%$), cooling time ($f_{\rm SCC} = 33\%$) and entropy ($f_{\rm CC} = 41\%$) in the Magneticum Pathfinder simulation \citep{dolag2015}. To compare to their result for the central cooling time and entropy we combine our SCC and WCC fractions. Our respective fractions are larger (lower) for $t_{\rm cool, 0}$ and $K_0$ ($n_{\rm e,0}$). However, the mean mass of the Magneticum cluster sample is unclear, and may differ substantially from TNG-Cluster.
In addition, as for Rhapsody-G above, the physical model as well as numerical simulation technique of Magneticum are both substantially different than in TNG. Cluster central properties and CC fractions can therefore inform our modeling of galaxy clusters. 

Observationally, considering the central cooling times of a sample of 30 nearby galaxy clusters from the REXCESS survey \citep[][spanning a mass range of $10^{14} \msun < M_{500} < 8 \times 10^{14} \msun$]{bohringer2007}, \cite{haarsma2010a} find a CC fraction of 33\%. However, this comparison is at face value only, as they compute $t_{\rm{cool}, 0}$ within a different radius of $0.03 r_{500}$ \citep[computation of $t_{\rm{cool}, 0}$ taken from ][]{croston2008} and define CCs as clusters with $t_{\rm{cool}, 0}<2$\,Gyr. For comparison, \cite{hudson2010} find a SCC (WCC) fraction of 44\% (28\%) in their flux-limited sample of 64 X-ray selected clusters using the same definition.

Both clusters samples are in fact X-ray selected samples, that suffer from CC bias. In particular, \cite{andrade-santos2017} found that cool-cores are over represented in X-ray selected samples compared to SZ selected surveys by a factor of 2.1–2.7 depending on metric (for \textit{Chandra} data). Adopting this correction factor reduces the inferred CC fractions to 16\% and 21\%, which are in reasonable agreement with our findings.

\begin{figure*}
    \centering
	\includegraphics[width=0.46\textwidth]{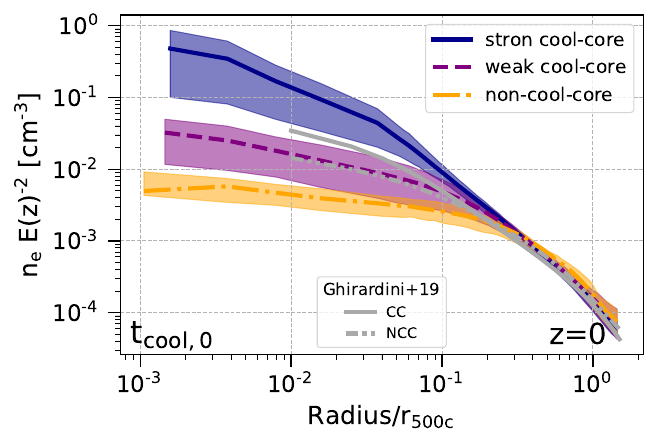}
    \includegraphics[width=0.46\textwidth]{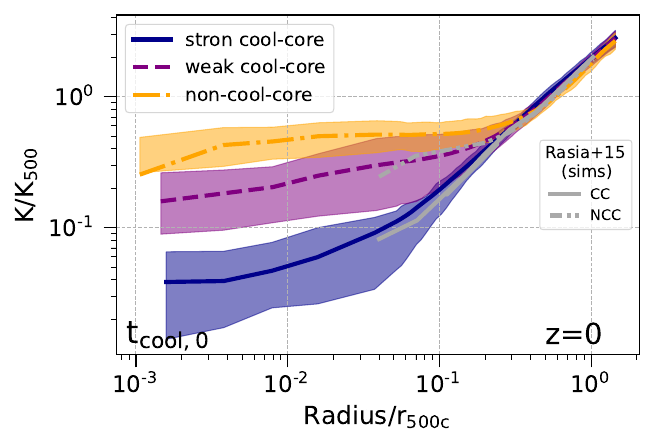}
	\includegraphics[width=0.46\textwidth]{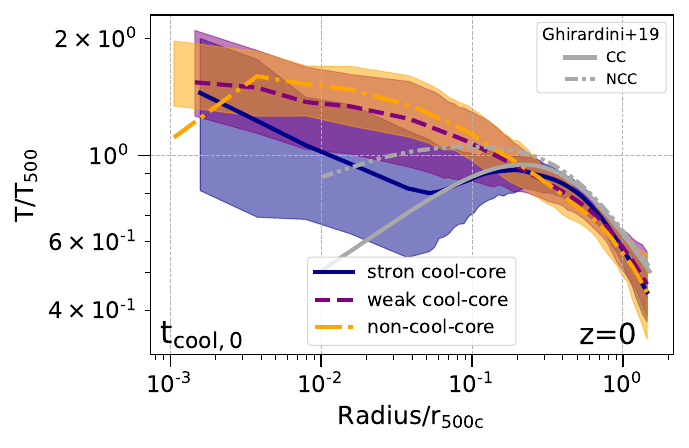}
	\includegraphics[width=0.46\textwidth]{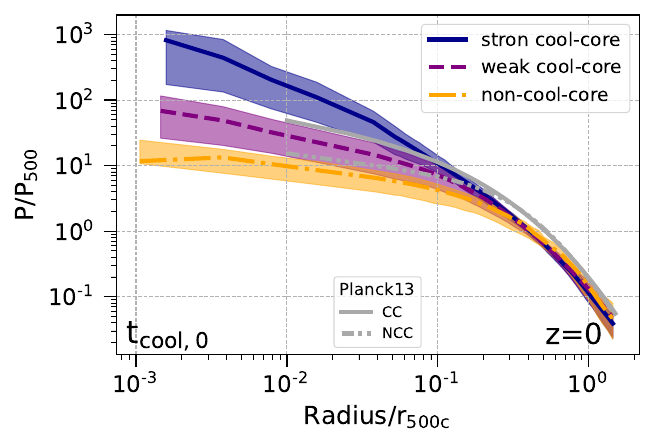} 
    \caption{Median profiles of four physical properties for SCC (dark blue, solid line), WCC (purple, dashed line) and NCC (orange, dashed dotted line) clusters for TNG-Cluster at $z=0$. From upper left to lower right the panels show: electron number density, entropy, temperature, and pressure. The filled regions show the 16-84 percentile band of the separate populations. The distinction in core states is made with our fiducial central cooling time definition. We compare to observations by \protect\cite{ghirardini2019} (electron number density and temperature), by the \protect\cite{planckcollaboration2013} (pressure), and to simulations by \protect\cite{rasia2015} (entropy). In general, the profiles for different populations are clearly separated and are offset in normalization while maintaining similar shape.}
    \label{fig:CCvsNCCprofiles}
\end{figure*}

Using central entropy, \cite{hudson2010} find a SCC fraction of 37\% and a WCC fraction of 34\%. While our SCC fraction is significantly smaller than the observed one, the combined fraction of SCCs and WCCs is in agreement with the combined observed sample. This suggests the split between SCCs and WCCs is not well-defined, i.e. that calculating fractions for each becomes very sensitive to the split threshold value.

\cite{haarsma2010a} also classify clusters using central electron number densities from \cite{croston2008} and find a SCC fraction of 33\% (for $n_{\rm e, 0}$ computed within 0.008$r_{500}$).

\cite{andrade-santos2017} use a sample of 164 clusters selected from the ESZ survey. They find (S)CC fractions based on central electron number density and cuspiness that are lower than ours ($f_{\rm SCC}= 39\pm 5 \%$ vs $52\pm 3\%$ and $f_{\rm CC}= 38\pm5\%$ vs $45\pm 4\%$). However, the average mass of the SZ selected sample is larger than the median mass of our total sample. The same work finds CC fractions based on the X-ray concentration parameter that are significantly larger than ours. To compare to the physical (scaled) concentration parameter we must combine the SCC and WCC fractions (consider a threshold of $C_{\rm{scaled}}>0.4$). Even so, fractions inferred from X-ray concentrations appear to be larger, by up to a factor of two, than in TNG-Cluster ($36\pm 5\%$ vs $18\pm 2\%$ [$C_{\rm phys}$], $28\pm 4 \%$ vs $18\pm 2\%$ [$C_{\rm scaled}$]). 

Summarizing, we find that our CC fractions are in tentative agreement with observed fractions. However, CC fractions strongly depend on the defining criteria, making careful comparisons essential. Although we try to compare to studies that employ similar CC criteria, differences remain. These include different apertures in which the central values are computed. An apples-to-apples comparison with data requires detailed forward modeling, and we reserve for future work such a careful comparison of samples and methodologies.

\subsection{CC vs NCC cluster structure (radial profiles)}\label{subsec_profilesCCvsNCC}

We return to cluster thermodynamical profiles, but now stack according to core state. Figure~\ref{fig:CCvsNCCprofiles} shows median profiles for electron number density (upper left), entropy (upper right), temperature (lower left), and pressure (lower right). The blue solid line shows stacked SCC profiles, the purple dashed line shows stacked WCC profiles, and the orange dash-dotted line shows combined profiles of NCCs. To separate the clusters in the three core states we use the central cooling time, our fiducial CC criterion.

On the upper left we show the electron number density profiles. They significantly differ in the core, by roughly two orders of magnitude. They begin to converge at a radius of $\sim 0.2 \rvir$ that separates the regions dominated by gravitational effects from the region where non-gravitational processes become important. Regardless of cool-core status, all three density profiles fall on top of each other at large radii. 

The SCC profile reaches the highest densities in the core ($n_{\rm e}(0.001\rvir) \sim 0.5 $\,cm$^{-3}$), while the NCC profile has the lowest core densities ($n_{\rm e}(0.001\rvir) \sim 0.004 $\,cm$^{-3}$). The stacked NCC profile becomes flat towards the center, increasing by less than a factor of two from $\sim 0.1 \rvir$ to $0.001 \rvir$. The stacked WCC profile has intermediate core densities. This finding is expected since CC clusters have a high central density and NCCs do not, by definition. Additionally, since we classify clusters based on central cooling time, SCCs have low $t_{\rm{cool}, 0}$ and thus high density or low temperature in the core. 

The upper right panel presents the entropy profiles. As before, they are well separated in the core and become similar at large radii. As expected, the SCC profiles have the smallest central entropy. The median NCC profile is again relatively flat in the center, with the highest values. 

For both, electron number density and entropy, the colored bands stating the uncertainty do not overlap for $r < 0.1 \rvir$. The three classes are well separated, implying that $n_{\rm e, 0}$ and $K_0$ are highly correlated with central cooling time. Additionally, measurements of these two parameter are rather effective in assigning clusters to a CC class (unlike temperature, see below), assuming central cooling time as the fiducial choice for classification.

In the lower left panel of Figure~\ref{fig:CCvsNCCprofiles} we show the stacked median temperature profiles. In contrast to the previous two cases, here the amplitude of separation is rather small and the shape of the profiles differ substantially. The NCC profile has the largest temperature across all radii. The NCC and WCC profiles monotonically increase until $0.004 \rvir$. The SCC profile also increases towards the center, but with a pronounced dip at $\sim 0.04 \rvir$. This dip makes the actual temperature of SCCs ($r \lesssim 0.5\times 10^{-2} \rvir$) cooler than WCCs and NCCs, although the SCC temperature recovers in the very center.

In the core there is no separation of the three profiles, and clusters have similar central temperatures at $r \lesssim 10^{-3} \rvir$ regardless of cool-core status. The differences in core temperature between the cluster states is small, causing the percentile bands to overlap. We speculate that the central temperatures of clusters, especially SCCs, are impacted by AGN feedback that heats their cores. We will return to the physical origin of these differences in central temperature with future work. 

The pressure profiles (lower right) show similar trends to the electron number density profiles. In the center the different cluster classes clearly separate. SCCs have the highest central pressure and NCCs the lowest. At larger radii, the three profiles approach each other. Note that in Figure~\ref{fig:CCvsNCCprofiles} we show only CC classifications based on central cooling time. The stacked profiles look similar when using other CC criteria. The amplitude of the separation varies with varying CC criteria, but the shape of the profiles stays roughly the same. 

Observationally, \citet{ghirardini2019} fit profiles to X-COP sample, split in CC and NCC clusters. They infer the profiles for the range $10^{-2}  < r/\rvir < 2.5$ and normalize to the virial value. In general, their trends with radius are in reasonable agreement with our profiles. 
However, the temperature profile for CC clusters continues to decrease for $r < 0.1 \rvir$, whereas our temperature profile increases again for smaller radii. While the separation for their electron number density, entropy, and pressure (both not explicitly shown) in the core is smaller compared to ours, their scatter in the temperature profile is comparable.

We also qualitatively compare to the pressure profiles presented by the \cite{planckcollaboration2013} (lower right panel). Their trends are in agreement with ours, although the separation in the core is smaller. This may be a consequence of splitting the sample into two classes instead of three.

We also compare to the results of the Rhapsody-G simulation \citep{hahn2017a}. They split their 9 clusters based on central entropy into CC and NCC samples. Their central electron number density and entropy profiles are well separated in the core, where the NCCs have a flatter profile with lower density/higher entropy in the core. At larger radii, the profiles of the different cluster populations approach each other. The temperature profiles also show similar trends, with comparable scatter, as ours. The SCC profiles exhibit a dip, while the NCCs have a flat profile in the center. Their SCC profiles do not feature the increase towards the very center that we see. However, the positions of the dip cannot directly be compared as \cite{hahn2017a} present radii in physical units, while we normalize to $\rvir$. 

Similarly, \cite{rasia2015} measure entropy profiles for the two cluster populations in their DIANOGA sample of 29 simulated clusters. The profiles are well separated in the core, and the stacked profile of the NCC population has a higher entropy in the core than the stacked SCC profile. Their profiles for the two cluster populations agree well with ours (see upper right panel in Figure~\ref{fig:CCvsNCCprofiles}). 

Summarizing, the stacked profiles of TNG-Cluster at $z=0$ when split into (N)CC populations are in reasonable agreement with inferences on observed profiles as well as the few available profiles from other numerical work.


\section{Evolution of CC/NCC Clusters with Redshift} \label{sec_zevo}

\begin{figure*}
    \centering
    \tabskip=0pt
    \valign{#\cr
      \hbox{\begin{subfigure}[b]{.66\textwidth}
        \centering
        \includegraphics[width=\textwidth]{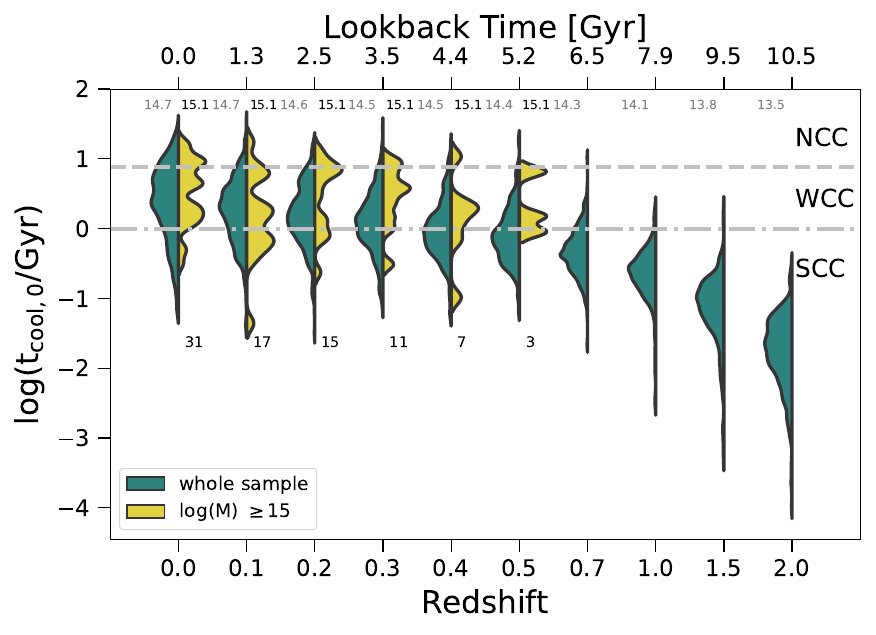}
        \end{subfigure}}\cr
      \noalign{\hfill}
      \hbox{\begin{subfigure}{.33\textwidth}
        \centering
        \includegraphics[width=\textwidth]{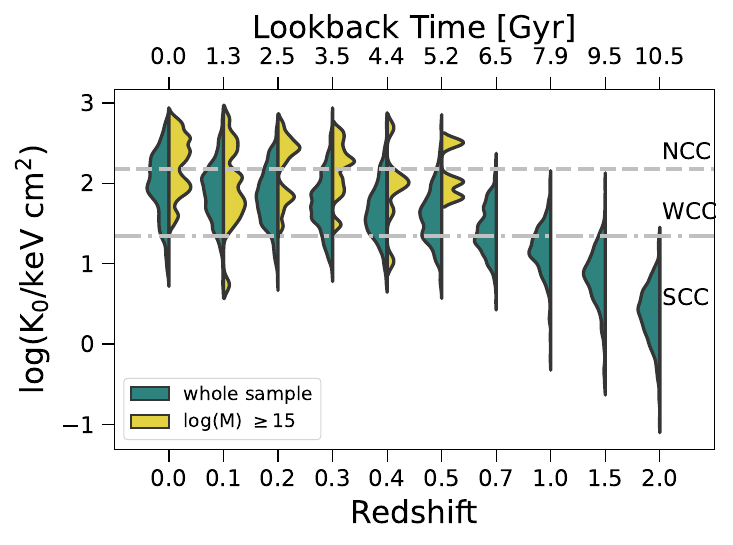}
        \end{subfigure}}
        \vfill
      \hbox{\begin{subfigure}{.33\textwidth}
        \centering
        \includegraphics[width=\textwidth]{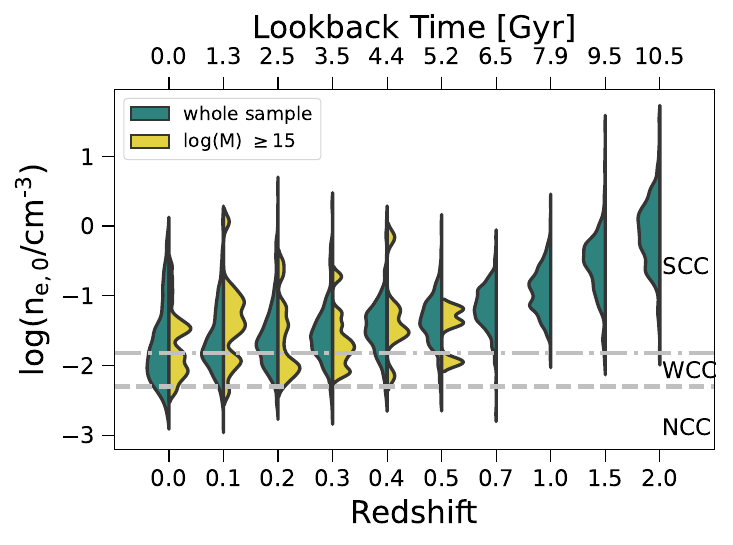}
        \end{subfigure}}\cr}
    \hbox{\begin{subfigure}{.33\textwidth}
        \centering
        \includegraphics[width=\textwidth]{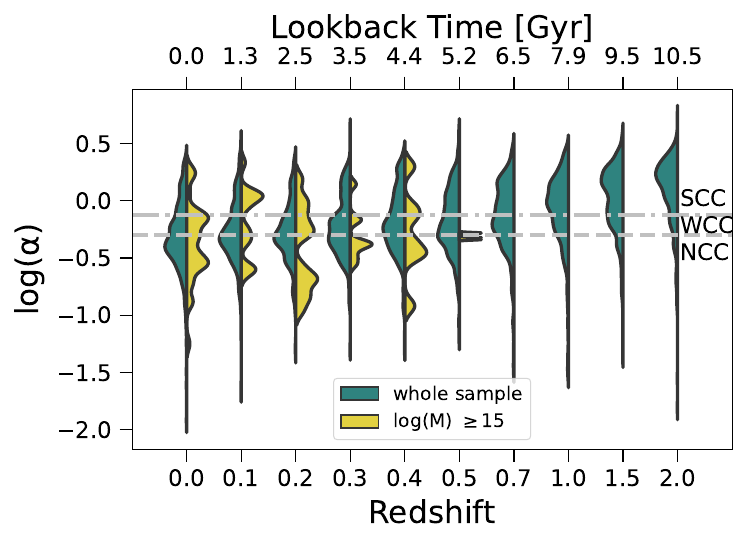}
        \end{subfigure}}
    \hbox{\begin{subfigure}{.33\textwidth}
        \centering
        \includegraphics[width=\textwidth]{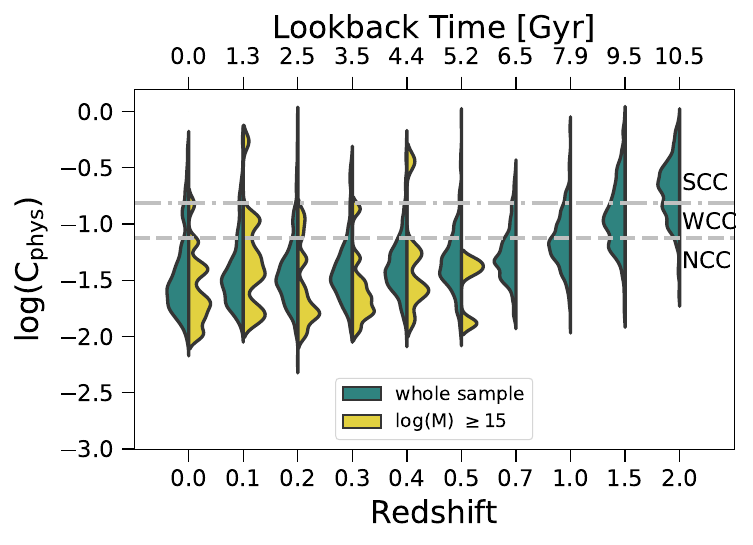}
        \end{subfigure}}
    \hbox{\begin{subfigure}{.33\textwidth}
        \centering
        \includegraphics[width=\textwidth]{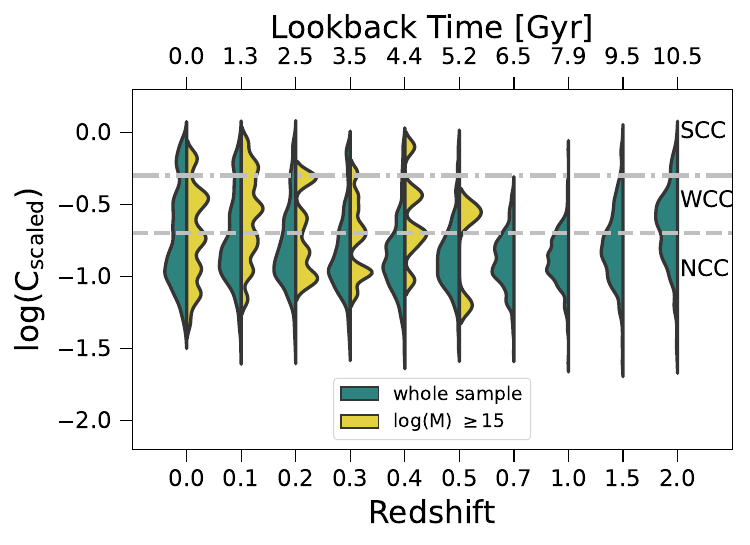}
        \end{subfigure}}
    \caption{Redshift evolution of the six central properties used to classify (non) cool-core clusters, from $z=0$ to $z=2$. These are, clockwise from the upper left: cooling time, entropy, electron number density, scaled X-ray concentration, physical X-ray concentration, and cuspiness. Each panel includes two sets of violins: the left (teal) shows the distributions for the entire TNG-Cluster sample, while the right (yellow) shows only high-mass with $\mvir > 10^{15}\,\msun$. In this latter case we probe evolution at approximately fixed halo mass, and the number in each bin is indicated. Horizontal gray lines mark the thresholds between NCC, WCC, and SCC clusters. Strong redshift evolution is due both to the evolving mass distribution of the sample, with lower mass clusters at earlier times, as well as physical evolution in central gas properties.}
    \label{fig:CCcritVsz}
\end{figure*}

After studying the cluster population at $z=0$, we now investigate how the core properties, the CC fractions and radial profiles evolve change with redshift. We qualitatively compare our findings to several observational samples up to $z=2$. In particular, cluster surveys based on selection via the Sunyaev-Zeldovich (SZ) effect provide samples free from the CC bias due to the redshift independence of the SZ signal \citep{birkinshaw1999}. This allows a less biased comparison to theoretical results from simulations than X-ray selected samples \citep{andrade-santos2017, barnes2018}.

We first consider the SPT-XVP sample of \cite{mcdonald2013}. It consists of 83 clusters in the mass range $\sim 2\times 10^{14} \msun/h_{70} < {\rm M}_{500} < 2\times 10^{15} \msun/h_{70}$ and a redshift range of $0.3 < z < 1.2$. The median redshift is $z \sim 0.7$. The clusters are SZ selected based on detection significance from SPT observations. All SPT-selected clusters in their work were then observed with \textit{Chandra} to characterize their physical properties via higher resolution X-ray data. The sample of \cite{mcdonald2014} of 80 clusters are selected from the SPT 2500 deg$^2$ survey and are similarly followed up with \textit{Chandra}. This sample spans a redshift range of $0.3 < z < 1.2$. Finally, \cite{mcdonald2017} presents a sample of the eight most massive galaxy clusters at $z>1.2$ from the SPT 2500 deg$^2$ survey - the SPT-Hiz sample. These clusters span a mass range of $\sim 2\times 10^{14} \msun < {\rm M}_{500} < 4\times 10^{14} \msun$ and a redshift range of $1.2 < z < 1.9$. For our comparison we consider X-ray peak centered CC fractions.

\cite{sanders2018} used almost the same sample of clusters compared to \cite{mcdonald2013}, differing only in a few halos. 
However, their analysis uses an alternate definition of the cluster center, and this leads to different results than the original analysis of \cite{mcdonald2013, mcdonald2014}. Contrary to the other studies, they also determine the central core properties within an aperture of fixed physical size of $\sim 10$\,kpc.

To study the evolution of cluster progenitors with time, \cite{ruppin2021} select high-z clusters that resemble progenitors of well-known low redshift clusters. 49 clusters of the sample by \cite{mcdonald2013} fulfil their criteria. Additionally, their sample include 18 clusters from the SPT 100 deg$^2$ catalog, for which \textit{Chandra} data are also available. The clusters are selected based on their mass growth rates, yielding a mass range of $6.4\times 10^{14} \msun < M_{500} < 1.3\times 10^{15} \msun$ at $z\sim0$. The 67 selected clusters span a redshift range of $0.25< z <1.3$. They use the X-ray centroid as a deprojection center and not the X-ray peak.

\subsection{CC criteria vs redshift}\label{ch:CCcritvsZ}

Figure~\ref{fig:CCcritVsz} presents the evolution of the six central cluster properties with redshift. At each redshift, a kernel density estimate (KDE) of the underlying distribution is used to represent the values in a violin plot. The left side (teal) of each violin shows the distribution of the total sample of 352 clusters, while the right side (yellow) shows the distribution of only high-mass clusters with $\mvir > 10^{15}\,\msun$. Below each violin we label the number of clusters in the high-mass bin and above each violin we label the median mass both in the full and high-mass sample. 

The main panel (upper left) shows the redshift evolution of the central cooling times. The left (teal) violins show a strong trend with redshift. The overall violin is shifted to lower $t_{\rm cool, 0}$ for higher redshift by two orders of magnitude; the mean values shift from $\log_{10}(t_{\rm cool, 0}/\rm{Gyr}) = 0.35$ at $z=0$ to $\log_{10}(t_{\rm cool, 0}/\rm{Gyr}) = -1.78$ at $z=2$. In contrast, the shape of the violins only change moderately: at all redshifts the distribution is continuous and exhibits no bimodality. The distribution is asymmetric and skewed to higher values for the lowest and highest redshifts, for intermediate redshifts the distribution is skewed to lower $t_{\rm cool, 0}$. Due to the strong shift of the overall violin to smaller $t_{\rm cool, 0}$, we classify nearly all clusters as SCCs for $z>0.7$. 

The median mass of the sample is changing strongly with redshift; $\log_{10} (\mvir/\msun) = 14.7$ at $z=0$, and $\log_{10} (\mvir/\msun) = 13.5$ at $z=2$. Thus, the strong redshift evolution is influenced both by the strong evolution of cluster mass and by the evolution of gas properties. The high-mass sample (yellow violin) keeps roughly the same median mass for the whole redshift range of $0 < z < 0.5$, and these distributions do not show significant redshift evolution. This may indicate that the pronounced redshift evolution of the full sample is strongly induced by the evolution of the cluster masses.

The change of central entropy (upper right panel) shows similar behavior: a strong redshift evolution towards lower entropy for the full sample, and only a moderate evolution of the high-mass sample.  The center right panel shows the evolution of the central electron number density. The trends are comparable to cooling time and entropy, and can explain the evolution of both.

Since the central density of the full sample rapidly increases for high redshift, we expect, by definition, the cooling time and entropy to decrease towards higher redshift. Since clusters are defined as overdensities relative to the critical density of the Universe, and given that this density increases with redshift, $n_{\rm e,0}$ will correspondingly increase with redshift at fixed halo mass. By defining $n_{\rm e, 0}$ relative to a fixed fraction of $\rvir$, the aperture also shrinks with increasing redshift, which, in turn, leads to an additional increase in $n_{\rm e, 0}$ \citep[see e.g.][]{barnes2018}. To remove this second effect it could be beneficial to compute $n_{\rm e, 0}$ within a fixed physical aperture of, for example, 10\,kpc. 

The evolution of the cuspiness parameter (lower left) is also towards more SCCs at higher redshift, but the trend is milder than in the three previous properties. The average value of the full sample evolves from $\log_{10} \alpha = -0.31 $ at $z=0$ to $\log_{10} \alpha \sim 0.07 $ at $z=2$. For the previous central properties the change in mass with redshift was the main driver of redshift evolution, in case of the cuspiness the effects of mass are less prominent. This is because the cuspiness is computed from the slope of the density profiles, but in Figure~\ref{fig:ProfilesNormedKTnePz0} we saw that the slope of the density profile at $0.04 \rvir$ does not vary strongly with mass. Since the mass trend is milder in cuspiness, it may be better suited for comparing samples across redshift.

The evolution of the physical concentration parameter (lower center) is comparable to the cuspiness parameter, but due to the imposed thresholds the clusters are mostly classified as NCCs at $z<0.7$. The average value of the full sample evolves from $\log_{10} \alpha = -1.43 $ at $z=0$ to $\log_{10} \alpha = -0.7 $ at $z=2$.  Assuming that the size of a cool core is fixed \citep{santos2008, mcdonald2017}, one can expect the physical concentration parameter to be less dependent on mass. If that is the case, $C_{\rm phys}$ is also a CC metric that is well suited for the study of the redshift evolution of cluster samples with a large scatter in mass.

\begin{figure*}
    \centering
	\includegraphics[width=0.46\textwidth]{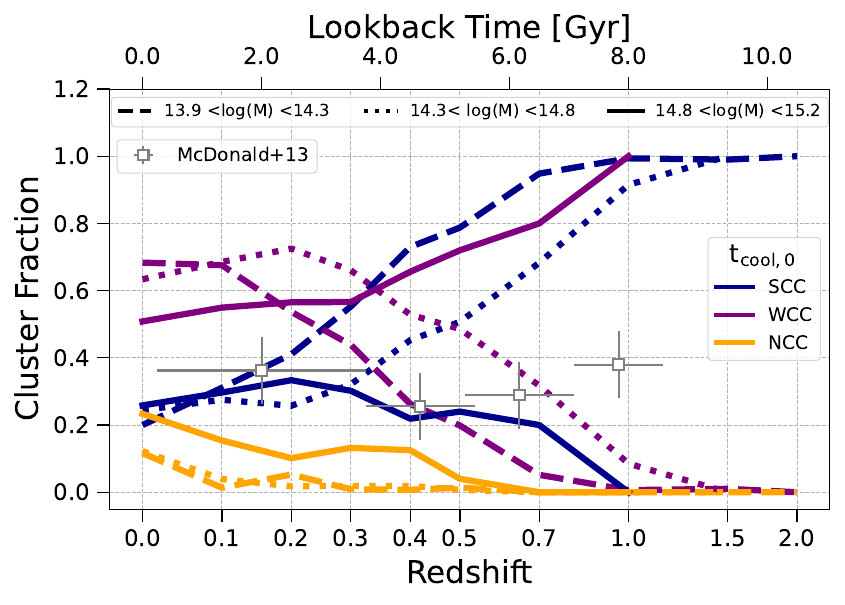}
	\includegraphics[width=0.46\textwidth]{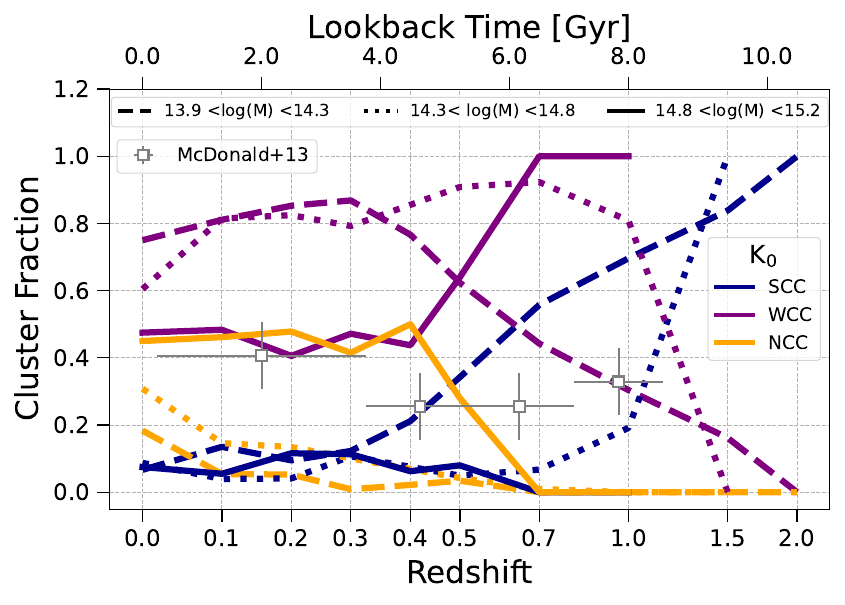}
	\includegraphics[width=0.46\textwidth]{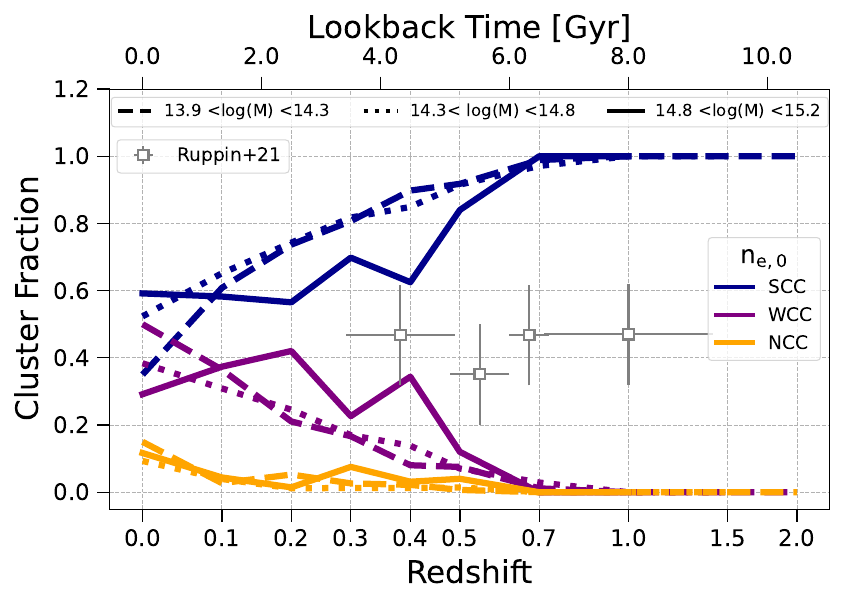}
	\includegraphics[width=0.46\textwidth]{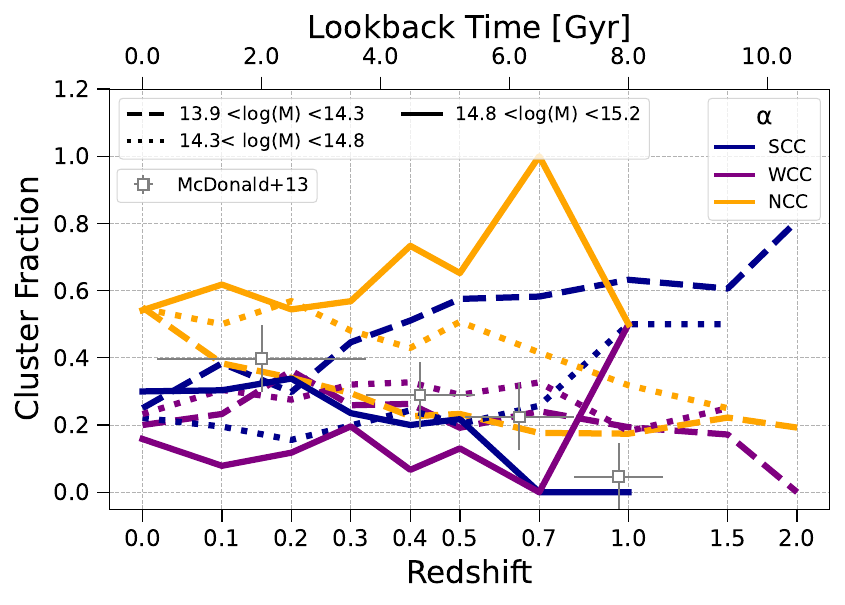}
    \includegraphics[width=0.46\textwidth]{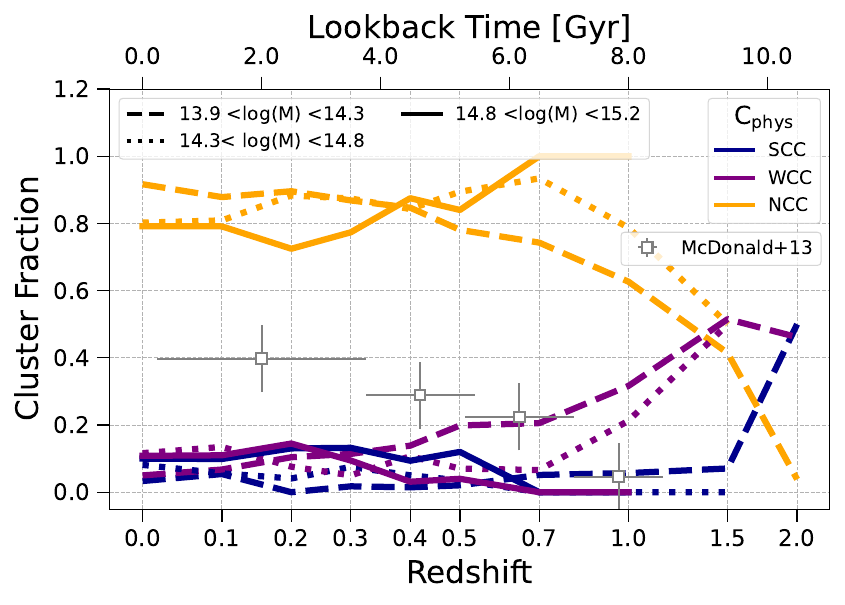}
    \includegraphics[width=0.46\textwidth]{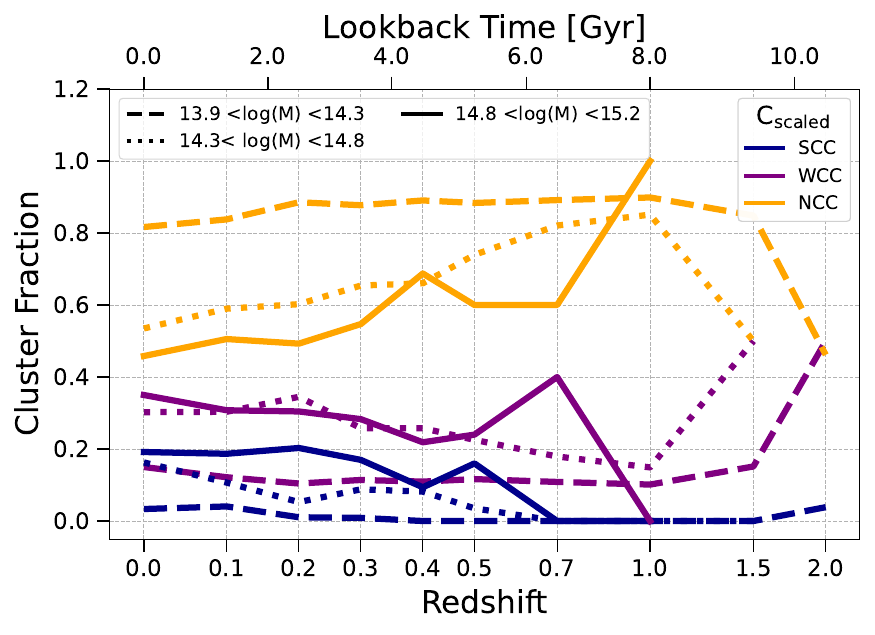}
    \caption{Fraction of strong cool-core (SCC; blue), weak cool-core (WCC; purple), and non-cool-core (NCC; orange) clusters in the TNG-Cluster simulation as a function of redshift. We separate clusters into low-mass (dashed lines), intermediate-mass (dotted lines), and high-mass bins (solid lines). The $\log_{10}({\rm M}/\mvir)$ thresholds for each mass bin are stated in the legend. We overplot the results for the SCC fraction from observations \protect \citep{mcdonald2013, ruppin2021} in five of the six panels.
    For $0<z \leq 0.7$ the observations should be compared to the high-mass bin (solid line) and at $z\sim 1$ to the intermediate-mass bin (dotted line). The comparisons are meant to be qualitative, at face value, only. The six panels show classification based on our six fiducial CC criteria, from top left to bottom right: cooling time, entropy, density, cuspiness, and the two X-ray concentration measures. Overall, the fraction of SCC clusters is higher at earlier times, and NCC clusters appear only at $z \lesssim 0.5$, although redshift trends depend on the adopted CC criterion.}
    \label{fig:fCCvsZ}
\end{figure*}

The evolution in scaled concentration parameter (lower right) is different from all previous CC metrics. From $z=0$ to $z=0.7$ the violins shift towards the NCC regime, i.e. towards lower concentrations. However, from $z=0.7$ to $z=2$ the violins are shifted towards the SCC regime. The trend with respect to $C_{\rm phys}$ can be explained by the differing definitions. Using fractions of $\rvir$ to define the concentration leads to a different sampling of cluster regions as a function of redshift. Since the same fraction of $\rvir$ in physical units is larger at low redshift, this will influence measurements of the scaled concentration parameter.

Observations of clusters up to $z=2$, reveal little to no evolution in CC criteria \citep{mcdonald2013, mcdonald2017, sanders2018, pascut2015}. In their sample spanning a redshift range of $0.3 <z<1.2$, \cite{mcdonald2013} find little evolution in $t_{\rm cool, 0}$ and $K_0$, and a weak evolution towards the NCC regime in $\alpha$ and $C_{\rm phys}$ with increasing redshift. In contrast to our sample, the average mass of their sample is similar across the whole redshift range. \cite{mcdonald2017} expands the previous study to redshifts $z\sim 1.9$, but find no measurable evolution. Focusing solely on the high-mass sample, which is less influenced by mass variations, our CC criteria trends are consistent with no evolution within their uncertainties.

\subsection{CC fraction vs redshift}\label{subsec_fccVSz}

Figure~\ref{fig:fCCvsZ} shows the redshift evolution of strong cool-core, weak cool-core, and non-cool-core cluster fractions. Each of the panels represents classification by our six CC criteria: central cooling time (upper left), central entropy (upper right), central electron number density (middle left), cuspiness (middle right), and X-ray concentration within physical (lower left) and scaled (lower right) apertures. The three colors indicate SCC fraction (blue), WCC fraction (purple), and NCC fraction (orange), while the three linestyles indicate different mass bins: low-mass (dashed), intermediate-mass (dotted), and high-mass halos (solid). Halos are placed into these mass bins, and classified into their cool-core status, based on their properties at each distinct redshift (we consider a $z=0$ based classification in Figure~\ref{fig:fCCvsZ_classz0}).

Considering the low-mass bin in the upper left panel (central cooling time) we can see that the SCC fraction increases with increasing redshift. At $z=1$ all clusters in that mass bin are classified as SCCs. Consequentially, the WCC and NCC fractions in the low-mass bin approach zero at that redshift. The NCC fraction is overall low for the central cooling time and the WCC fraction decays from $\sim 70\% $ at $z=0$ to zero at $z=1$. The redshift trend in the intermediate-mass bin is similar, differing in the redshift at which the SCC fraction approaches 100\% ($z=1.5$). Overall, the SCC/WCC fractions in this mass bin drop faster with decreasing redshift than those for the low-mass bin. This implies that intermediate mass clusters are transformed faster to NCCs than low-mass clusters. This may arise because more massive clusters experience disruptive mergers earlier, and thus their cool cores are destroyed earlier. 

In contrast, in the high-mass bin the SCC fraction is increasing with decreasing redshift and the WCC fraction is decreasing. At $z=1$ all clusters in the high-mass bin are classified as WCCs. The high-mass bin of SCCs contains only 7 ($z=0.4$), 6 ($z=0.5$), and 1 ($z=0.7$) clusters at the stated redshifts. This is also due to the imposed CC thresholds. In Sec.~\ref{ch:CCcritDef} we motivated the thresholds for the cooling time at $z=0$, however these arguments should be adjusted for high-redshift clusters. The evolution of the high-mass NCC fraction is similar to the other mass bins, although the overall fractions are larger.

The central entropy (upper right) and the cuspiness parameter (center right) show similar redshift and mass trends compared to the central cooling time. However, the absolute values of the cluster fractions are different. For example, the cuspiness-derived NCC fraction is the dominant class for all mass bins and redshifts. The transformation of SCCs into WCCs/NCCs is more rapid when fractions are computed using the central entropy. 

When classifying based on $n_{\rm e, 0}$ (center left) all mass bins show the same redshift trend, which is also similar to the redshift trend of the low-/intermediate-mass bin using central cooling time. The fractions using the physical concentration parameter (lower left) have similar mass trends compared to $n_{\rm e, 0}$. However, most of the clusters are classified as NCCs across all mass bins and all redshifts. Nonetheless, the NCC fraction is decreasing for increasing redshift.

The redshift trend of the fractions using the scaled concentration parameter (lower right) is different. Most of the clusters are classified as NCCs in all mass bins across all redshifts, but the NCC fraction is increasing with increasing redshift. In contrast, the SCC/WCC fractions are mildly increasing with decreasing redshift for all mass bins. These trends are opposite compared to all other CC metrics, but this is a consequence of the definition (for a discussion see Sec.~\ref{ch:CCcritvsZ}).

Summarizing, the SCC fraction is larger at higher redshift, and so the NCC fraction decreases towards higher redshifts. However, the absolute values of these fractions strongly vary with the CC criterion. Broadly speaking, the SCC fraction is larger in the low-mass bin compared to the high-mass bin. For most CC criteria, all mass bins evolve similarly with redshift.

In the following we compare to several observational works, whose important aspects are described in Sec.~\ref{sec_zevo}. The comparison is meant to be qualitative, i.e. without having replicated or mocked the measurement procedures of the observations. 

Given the \cite{mcdonald2013} sample, we must compare the observations in the redshift range $0<z\leq0.7$ to our high-mass bin, and for higher redshifts $z> 0.7$ to our intermediate-mass bin. The median mass of their sample in the last redshift bin is $\log_{10}({\rm M}/10^{14}\msun) \sim 14.56 $, while our median mass in the intermediate-mass bin at $z=1$ is $\log_{10}({\rm M}/10^{14}\msun) \sim 14.564 $. That work finds cool-core fractions, defined using central cooling or central entropy, have no strong evolution across their whole redshift range. In their four redshift bins, the SCC fractions fluctuate between $0.2 < f_{\rm SCC} < 0.4$ with no clear trend. Our two corresponding SCC fractions (solid line, upper two panels) are in broad agreement with the observed ones for $0<z\leq0.7$. Our SCC fraction, based on entropy, is lower, but \cite{mcdonald2013} uses a higher entropy threshold to define the CC fraction. At $z=1$, our SCC fraction is larger than (comparable to) the observed one for a classification based on central cooling time (entropy), based on our intermediate-mass bin (dotted line).

In the cool-core fractions based on cuspiness (center right) and physical concentration parameter (lower left), \cite{mcdonald2013} infer a decrease in CC fraction from $\sim 0.4$ at $z=0$ to $\sim 0.1$ at $z=1.2$. For cuspiness, our SCC fraction in the high-mass bin matches the observations well. For physical concentration the observed trend matches the redshift evolution of the high-mass bin, but at all redshifts our SCC fractions are smaller. However, we note that \cite{mcdonald2013} adjust the thresholds for $K_0$ and $C_{\rm phys}$ to reduce the scatter in CC fraction between different CC metrics in the low-redshift bin.

\cite{ruppin2021} finds a $n_{\rm e, 0}$-based SCC fraction that is compatible with no redshift evolution for $0.3 < z < 1.3$ ($f_{\rm SCC} \sim 0.5$). Since they aim to investigate clusters that resemble the progenitors of well-known low-redshift clusters, we should compare to the CC fractions of the high-mass bin based on the $z=0$ classification from Figure~\ref{fig:fCCvsZ_classz0}, but these are largely consistent with Figure~\ref{fig:fCCvsZ} in this case. In that redshift range our SCC fraction evolves from $\sim 0.75 $ at $z=0$ to $\sim 1$ at $z>1$.

In earlier simulations, \cite{burns2007} find a slightly increasing CC fraction with increasing redshift. Their sample covers a redshift range of $0<z<1$. However, the within the uncertainties the trend is also compatible with no redshift evolution. The clusters are assigned to cool-core classes based on the central temperature drop. On the other hand, \cite{planelles2009a} infer a redshift trend in their simulated cluster sample that is compatible with no evolution in the redshift range $0<z<1$, but find a decrease in CCs for higher redshifts $1<z<2$. The clusters are classified using the central temperature drop. In this simulation the effects of mergers on the cool-core status are more decisive, as they do not include feedback.

\begin{figure*}
    \centering
	\includegraphics[width=0.48\textwidth]{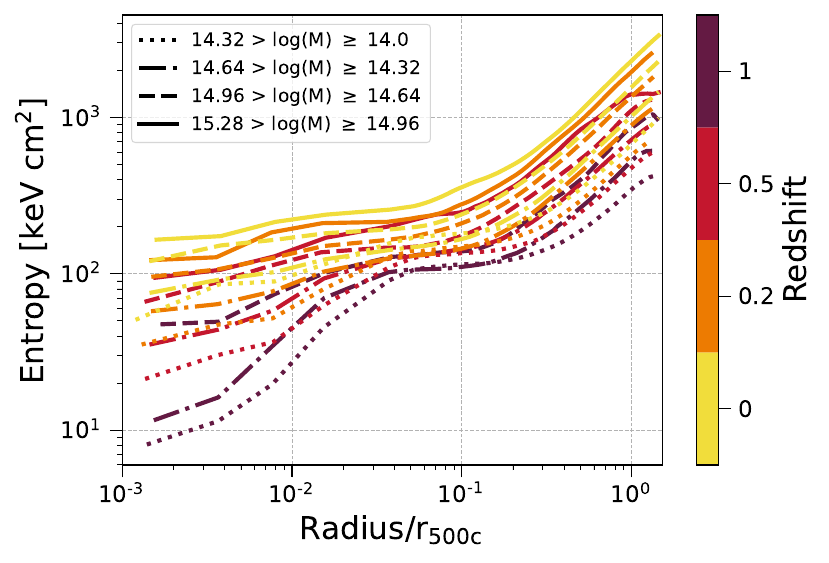}
    \includegraphics[width=0.48\textwidth]{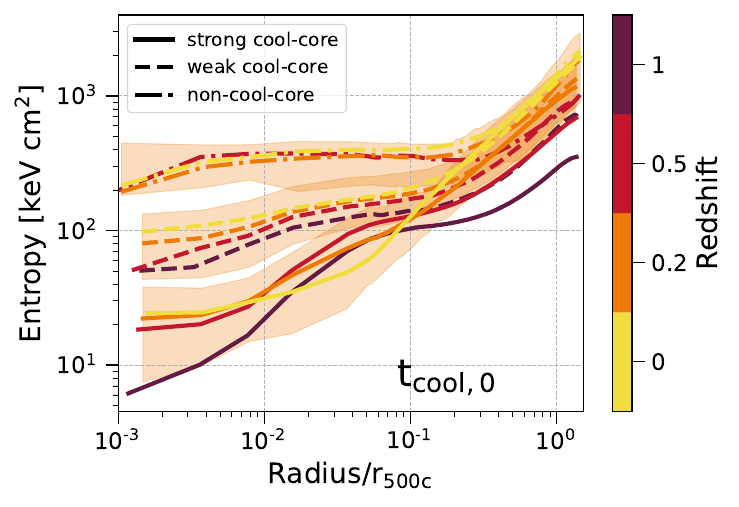}
    \includegraphics[width=0.48\textwidth]{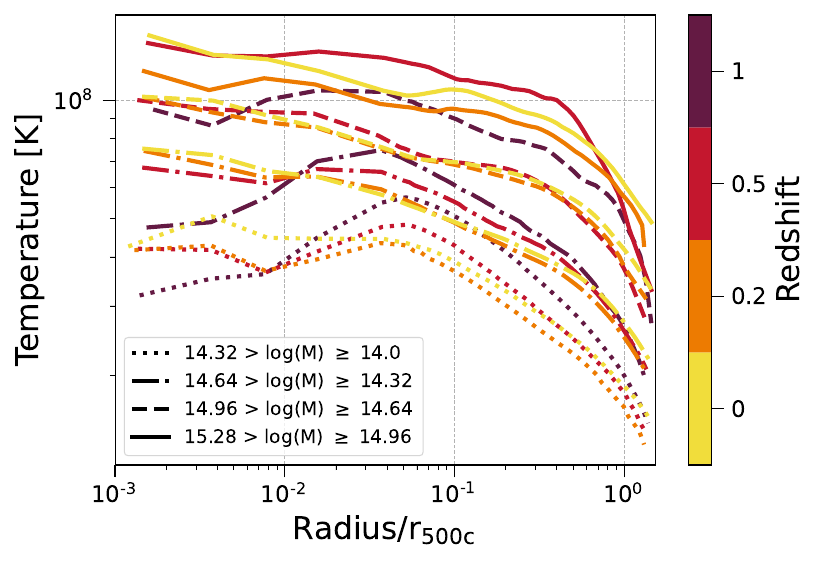}
    \includegraphics[width=0.48\textwidth]{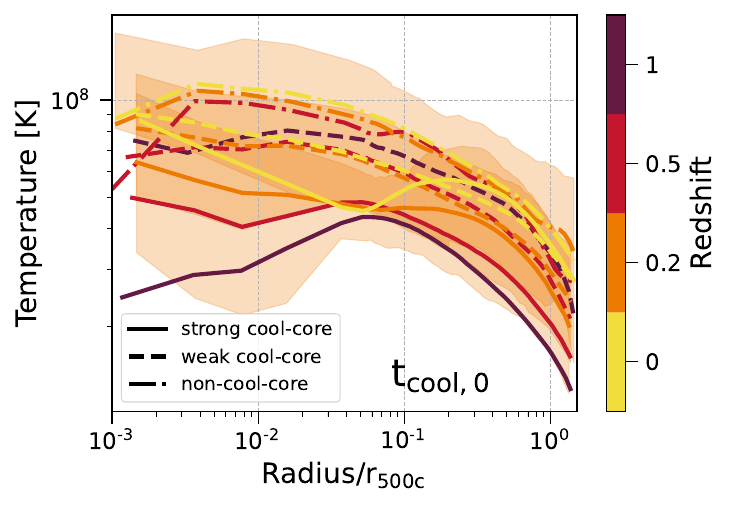}
    \caption{The redshift evolution of cluster entropy (top) and temperature (bottom) profiles. The left panels stack the full TNG-Cluster sample based on four halo mass bins ($\log_{10}({\rm M}/\mvir)$ ranges in legends), while the right panels separate the full sample into its SCC, WCC, and NCC sub-populations. In both cases, we bin in mass or classify cool-core state at each of $z=0$, $z=0.2$, $z=0.5$, and $z=1$ separately. The entropy profiles shift downwards with increasing redshift and decreasing halo mass, while maintaining a similar shape. The temperature profiles strongly depend on mass and weakly on the redshift. At all redshifts the entropy profiles for CCs and NCCs are well separated in the core, while the temperature profiles are not, and show a large scatter.}
    \label{fig:ProfilesVsRedshift}
\end{figure*}

The work on TNG300 led to similar results as ours \citep{barnes2018}. In their high-mass sample, the CC fractions based on central cooling time, density, and entropy are slightly steeper, or possibly in tentative agreement, with the observed trends. The fractions based on cuspiness and physical concentration are compatible with no redshift evolution, but in case of $C_{\rm phys}$ the normalization is offset.

\subsection{Redshift evolution of cluster profiles}\label{subsec_profilesVSz}

Figure~\ref{fig:ProfilesVsRedshift} shows the redshift evolution of cluster profiles, where we focus on entropy (top panels) and temperature (bottom panels). The qualitative trends are similar for density, pressure, and cooling time. The left panels focus on evolution at fixed halo mass. At each of four redshifts $z = \{0, 0.2, 0.5, 1.0\}$ (different colors) we select clusters in four halo mass bins (different linestyles). Mass bins are chosen to split the full sample in bins with $\sim 0.3$ dex width. In the right panels we classify clusters into SCC, WCC, and NCC subsets (different linestyles) according to their central cooling times at each redshift.\footnote{The apparent redshift evolution differs if we instead classify clusters at $z=0$, based on their $z=0$ central cooling time, and consider the stacked profiles of those progenitor subsets (see Appendix \ref{sec_appendix_zevo})}

Considering the upper left panel, we find that the entropy profiles in all mass and redshift bins decrease towards the center. The slope is steeper (mildest) for the clusters at $z=1$ ($z=0$). Nonetheless, the shape of the entropy profiles is similar for all cases. At a fixed mass bin the entropy is decreasing for increasing redshift, with the evolution strongest in the core. This is expected as density increases towards higher redshift at fixed mass. Additionally, the scaled radius at higher redshift probes smaller regions, causing an even higher density/lower entropy in the center. At fixed redshift the entropy increases with increasing mass. This is likely due to the strong mass trends of the temperature profiles (see below), whereas the mass trend of density is only moderate. The difference between mass bins is larger in the core and for larger redshifts. In other words, clusters become more uniform with time, in that sense that the low-mass clusters approach the high-mass clusters. This likely reflects hierarchical assembly: at $z\sim 1$ the massive systems already experienced disruptive mergers, while less massive systems are still evolving. By $z=0$ even the less massive halos underwent disruptive merging processes causing their central entropy to rise. 

At fixed mass, temperature (lower left) roughly decreases in the core towards higher redshift. However, at higher halo masses this trend is only marginal. It also inverts in the outskirts, although the profiles hardly change with redshift in the outskirts at fixed mass. At fixed redshift, the temperature increases for all radii by a similar amplitude with increasing mass. It is evident that changes in cluster mass have a more profound effect on the temperature profiles than changes in redshift, especially in the cluster outskirts. In the center, however, this effect is superimposed with other processes, such as AGN feedback. 

The temperature profiles for clusters in the high-mass bin (solid line) increase towards the center. In the second most massive bin (dashed line) this is also true, except for the $z=1$ profile where a small dip in the central temperature occurs. This dip is intensified in the lower intermediate-mass bin (dash-dotted line) relative to the others. In the low-mass bin (dotted line) the profiles at all redshifts (except $z=0$) feature such a dip. This implies that heating processes from e.g. AGNs are more important at lower redshifts and affect the centers of less-massive clusters more strongly. 

Next, we consider the redshift evolution of the internal structure of the (N)CC populations (right column in Figure~\ref{fig:ProfilesVsRedshift}). The entropy profiles (upper right) are for all redshifts well separated in CC states in the central region. There is no redshift evolution in the NCC entropy profiles. WCC profiles evolve moderately in the core towards smaller entropy with increasing redshift. For SCC clusters this trends is even more pronounced. At the outskirts the profiles for the three different populations approach each other for fixed redshift. At lower redshifts, there is less separation between the three profiles, and a shift towards higher entropy values. The shape of the profiles of each CC class stays similar across all redshifts with the central dip increasing with redshift. Finally, as the samples are defined at every redshift, clusters can change populations from CC to NCC and vice versa.

The temperature profiles (lower right) are not well separated by CC state. This is also evident from the orange band, which gives the 16-84 percentile region for the profiles at $z=0.2$. We show this band as representative in all cases, to avoid overcrowding. Temperature separation gets larger with increasing redshift. This may be induced by the stronger variation in median mass in the SCC profiles (see above) and by the strong mass dependence of the temperature profiles. Impressively, the NCC profiles show no clear trend with redshift and keep the same shape for $0<z<2$. There is moderate redshift evolution in the WCC profiles, and at higher redshift a shallow dip in the core of the profiles appears. The SCC profiles evolve strongly in shape with redshift: at $z=0$, a strong dip occurs at $0.04 \rvir$ and the temperature increases again towards the center, while at $z\ge1$ the SCC profiles decrease towards the center from this point. Using different CC criteria leads to similar results and redshift evolution. 

We compare the TNG-Cluster redshift evolution to several observational works. \cite{mcdonald2013} shows entropy profiles for CC vs NCC clusters in three redshift bins ($z<0.1$, $0.3<z<0.75$, and $0.75 < z < 1.1$) classified based on $K_0$. The radial entropy profiles match ours reasonably well. At $\rvir$ they find decreasing entropy with increasing redshift, matching the TNG-Cluster expectation. At their innermost radius of $0.01r_{\rm 500}$ they observe no evolution. This is in broad agreement with our profiles, especially when we classify the clusters based on $K_0$ (not explicitly shown). In contrast, they find substantial evolution in gas density profiles. The density of SCCs increases towards lower redshift in the core, while the core values of NCCs stay roughly constant. These findings also match the behavior of our electron number density profiles well (not shown).

\cite{mcdonald2014} presents CC and NCC profiles for two different redshifts ($z \sim 0.5$ and $z \sim 0.8$) for normalized entropy and temperature. They split clusters based on cuspiness, classifying halos above the mean as CCs. The entropy at $0.01\rvir$ does not evolve for the CC population but increases for the NCC population. This is different from our findings, even when we adopt cuspiness as our CC criterion. The temperature of their CC profiles decreases with increasing redshift at $0.01r_{500}$, which is compatible with our findings. However, their NCC profiles do not evolve with redshift, which also differs from TNG-Cluster. However, they use a different threshold in cuspiness and the show entropy/temperature profiles normalized to $K_{500}$/$T_{500}$, which makes the comparison non-trivial.

Physical evolution of cluster properties, from the cores to their outskirts, clearly results from the interplay of different physical processes. These include: the evolution of the ICM attributed to the expansion of the Universe and structure growth, cooling processes within the ICM, and the heating of the ICM via AGN feedback or cluster mergers. The statistics of TNG-Cluster and the initial quantitative census of this work lay the groundwork for a detailed investigation into the dominant physics driving cluster evolution and, especially, cluster core transformation across time.


\section{Conclusions} \label{sec_conclusions}

In this paper we study the intracluster medium (ICM) of massive galaxy clusters using the new TNG-Cluster simulation, a suite of 352 high-mass galaxy clusters spanning a halo mass range of $10^{14} < \mvir /\msun < 2 \times 10^{15}$ at $z=0$. We focus on the central thermodynamical properties of the ICM, contrasting the physical properties, relative fractions, and evolution of cool-core (CC) versus non-cool-core (NCC) cluster populations. The key findings of this work are:

\begin{itemize}
    \item TNG-Cluster produces a large diversity of cluster core morphologies, in terms of density, temperature, entropy, and cooling time (Figs.~\ref{fig:CoreGallery_ne}~to~\ref{fig:CoreGallery_tcool}). Morphological features resembling well-known observed galaxy clusters exist. Radial profiles of density, temperature, entropy, and pressure at $z=0$ are in reasonable agreement with observed profiles (Figure~\ref{fig:ProfilesNormedKTnePz0}). 
    \item We consider six common cool-core criteria: central cooling time, entropy, electron density, the density profile slope, and two X-ray concentration parameters. All six distributions are unimodal, and no clear bimodalities are present. As a result, CCs and NCCs represent the two extremes of each distribution, and threshold values commonly used to classify CC status do not correspond to distinct features in our distributions. The simulated distributions are in broad agreement with observational findings from SZ-selected samples (Figure~\ref{fig:HistCCcrit}).
    \item Cluster core properties have diverse trends with halo mass. Some evolve towards the strong cool-core (SCC) regime as mass increases, some evolve towards the non-cool-core (NCC) regime, while others have no mass trend (Figure~\ref{fig:CCcritVsM500}). As a result, the abundance of (N)CCs -- overall, and as a function of mass -- strongly depends on the adopted CC criterion.
    \item The SCC fraction increases with higher halo mass for three CC metrics (cooling time, density, and scaled concentration), while for the other three CC criteria the CC fraction is consistent with no mass trend (entropy, cuspiness, and physical concentration). The NCC fraction rises with increasing mass (central cooling time and entropy), decreases (density and cuspiness), or shows no trend (both X-ray concentration parameters; Figure~\ref{fig:fCCVsM500}). 
    \item Using our fiducial CC criterion, central cooling time, we find a strong cool-core fraction of $f_{\rm SCC} = 24\pm 2\%$, a weak cool-core fraction of $f_{\rm WCC} = 60\pm 3\%$, and a non-cool-core fraction of $f_{\rm NCC} = 16\pm 2\%$ at $z=0$. The fractions vary considerably between CC criteria (see Tab.~\ref{tab:fCC_wholeSample}). TNG-Cluster cool-core fractions at $z=0$ are in broad agreement with observations (Figure~\ref{fig:fCCvsPaper}). 
    \item Radial profiles of electron number density, entropy, and pressure split by cool-core status are well separated in the core, and approach each other at $\rvir$ (Figure~\ref{fig:CCvsNCCprofiles}). Similarly splitting the temperature profiles does not produce a strong separation of the three classes, but the median SCC temperature profile exhibits a pronounced dip in the core.
    \item If we assume that the central cooling time is, from a theoretical perspective, the optimal parameter to distinguish core state, this implies that central density and entropy are highly effective observable probes of CC state.
    \item Three core properties (central cooling time, entropy, and density) have a strong redshift dependence towards the SCC regime from $0\leq z\leq2$ (Figure~\ref{fig:CCcritVsz}). This trend is due to physical evolution of central ICM properties, but also due to the lower mean mass of the TNG-Cluster sample at higher redshifts. The decreasing (physical) aperture, and the background evolution, lead to higher core densities at earlier times. These induce lower central cooling times, and lower entropy. The redshift trends of cuspiness and X-ray concentration are also towards the SCC regime, but less steep.
    \item Generally speaking, SCC (NCC) fractions are larger (smaller) at higher redshift. However, the actual fractions strongly vary with the adopted CC criterion. On average, SCC fractions are larger for lower mass versus high-mass clusters. For three CC criteria (central cooling time, entropy, and density) the slope of the CC fraction as a function of redshift may be steeper than in observations. On the other hand, the redshift trends for cuspiness, and X-ray concentration, are in tentative agreement (Figure~\ref{fig:fCCvsZ}). Sample matching and observational selection effects complicate these comparisons.
    \item Radial profiles of ICM properties also evolve with redshift. Cluster entropy profiles across $0\leq z\leq1$ shift towards lower entropy at earlier times (Figure~\ref{fig:ProfilesVsRedshift}), as also suggested by data. On the other hand, we find that mass trends dominate any redshift evolution for temperature profiles.
\end{itemize}

Our census of TNG-Cluster central ICM properties, and the evaluation of common cool-core criteria, supports the notion that cool-core (CC) and non-cool-core (NCC) clusters are not physically distinct classes. Instead, they represent the two extremes of otherwise continuous and singly peaked distributions of central ICM properties. Indeed, TNG-Cluster suggests that CCs and NCCs reflect a continuous metamorphosis of cluster core states.

What physical mechanisms drive the transformation from a strong cool-core into a non-cool-core? Can NCCs transition back to CCs? Future work can investigate the underlying mechanisms -- the how, and why -- using the TNG-Cluster simulation, as it allows us to study the impact of intertwined astrophysical processes including mergers, AGN feedback, and cooling, in realistic clusters and realistic cluster core environments.

\section*{Data Availability}

The IllustrisTNG simulations themselves are publicly available and accessible at \url{www.tng-project.org/data} \citep{nelson2019a}, where the TNG-Cluster simulation will also be made public in the near future. Data directly related to this publication is available on request from the corresponding authors.

\section*{Acknowledgements}

KL acknowledges funding from the Hector Fellow Academy through a Research Career Development Award. DN acknowledges funding from the Deutsche Forschungsgemeinschaft (DFG) through an Emmy Noether Research Group (grant number NE 2441/1-1). Moreover, this work is co-funded by the European Union (ERC, COSMIC-KEY, 101087822, PI: Pillepich). NT acknowledges that the material is based upon work supported by NASA under award number 80GSFC21M0002. KL and ER are Fellows of the International Max Planck Research School for Astronomy and Cosmic Physics at the University of Heidelberg (IMPRS-HD). The TNG-Cluster simulation has been executed on several machines: with compute time awarded under the TNG-Cluster project on the HoreKa supercomputer, funded by the Ministry of Science, Research and the Arts Baden-Württemberg and by the Federal Ministry of Education and Research. The bwForCluster Helix supercomputer, supported by the state of Baden-Württemberg through bwHPC and the German Research Foundation (DFG) through grant INST 35/1597-1 FUGG. The Vera cluster of the Max Planck Institute for Astronomy (MPIA), as well as the Cobra and Raven clusters, all three operated by the Max Planck Computational Data Facility (MPCDF). The BinAC cluster, supported by the High Performance and Cloud Computing Group at the Zentrum für Datenverarbeitung of the University of Tübingen, the state of Baden-Württemberg through bwHPC and the German Research Foundation (DFG) through grant no INST 37/935-1 FUGG. This analysis has been carried out on the VERA supercomputer of the Max Planck Institute for Astronomy (MPIA), operated by the Max Planck Computational Data Facility (MPCDF).

\bibliographystyle{aa}
\bibliography{refs}


\clearpage
\begin{appendix}
\section{Redshift Evolution of Progenitors} \label{sec_appendix_zevo}

\begin{figure}
    \centering
	\includegraphics[width=0.48\textwidth]{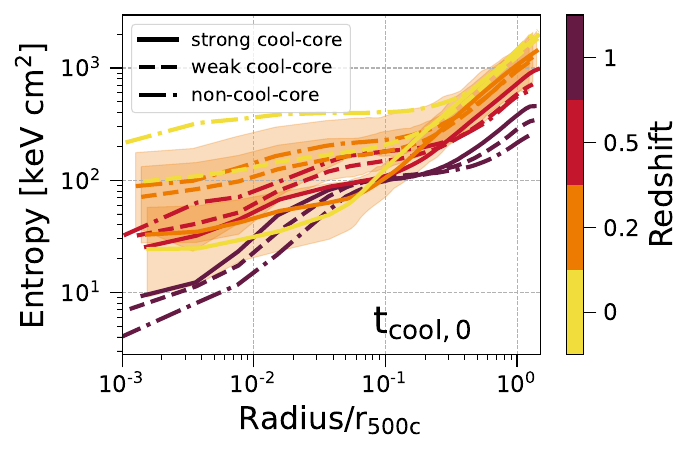}
    \caption{Evolution of cluster entropy profiles with redshift, as a function of SCC, WCC, or NCC classification. Analogous to Figure \ref{fig:ProfilesVsRedshift}, except that here we classify clusters at $z=0$, based on their $z=0$ central cooling time, instead of reclassifying clusters at each redshift.}
    \label{fig:ProfilesVsRedshiftz0class}
\end{figure} 

\begin{figure*}
    \centering
	\includegraphics[width=0.42\textwidth]{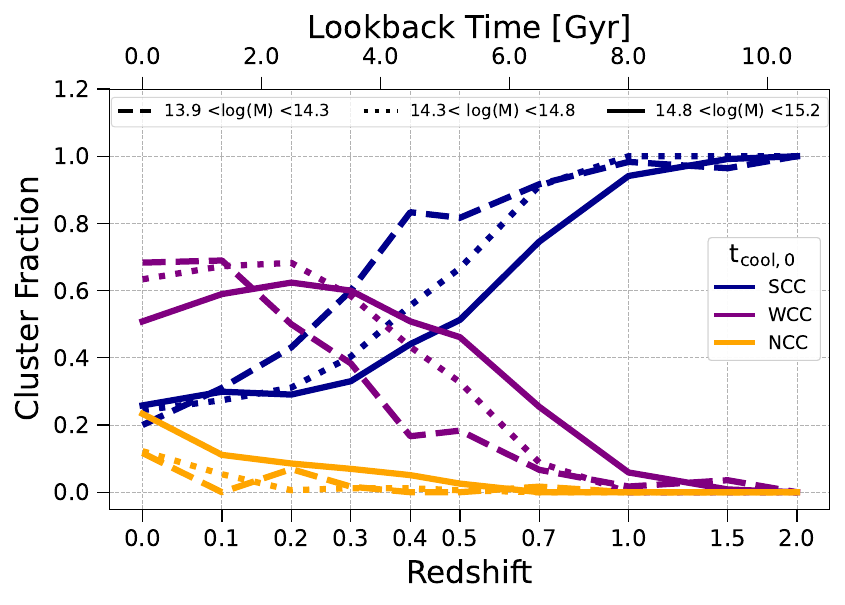}
	\includegraphics[width=0.42\textwidth]{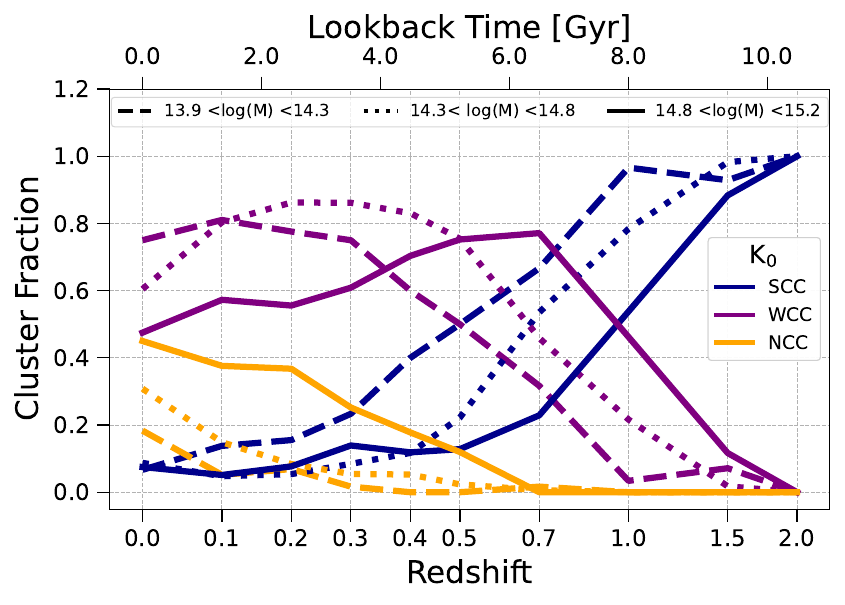}
	\includegraphics[width=0.42\textwidth]{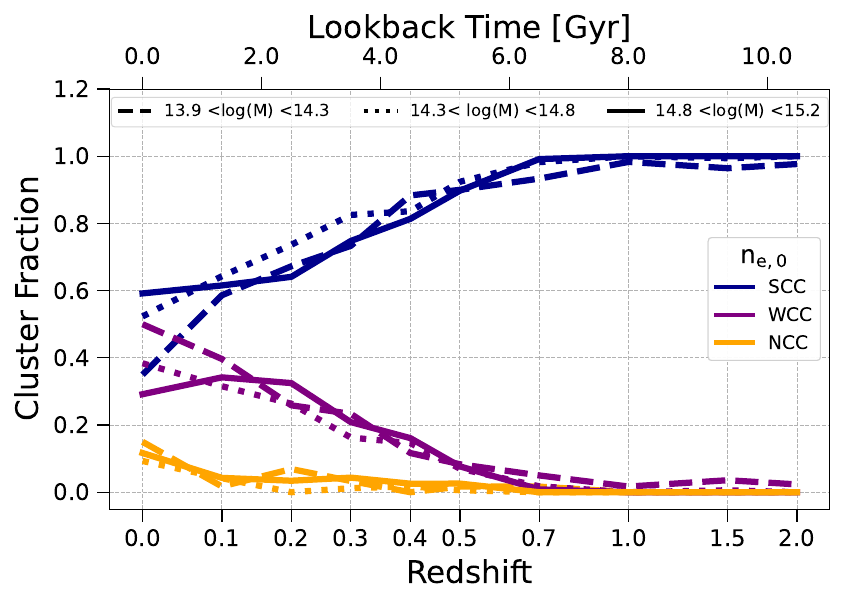}
	\includegraphics[width=0.42\textwidth]{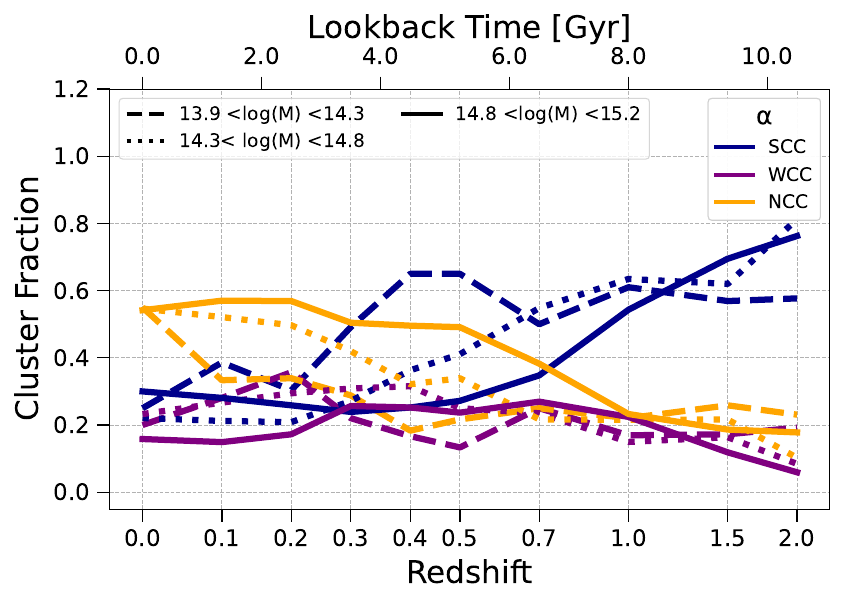}
    \includegraphics[width=0.42\textwidth]{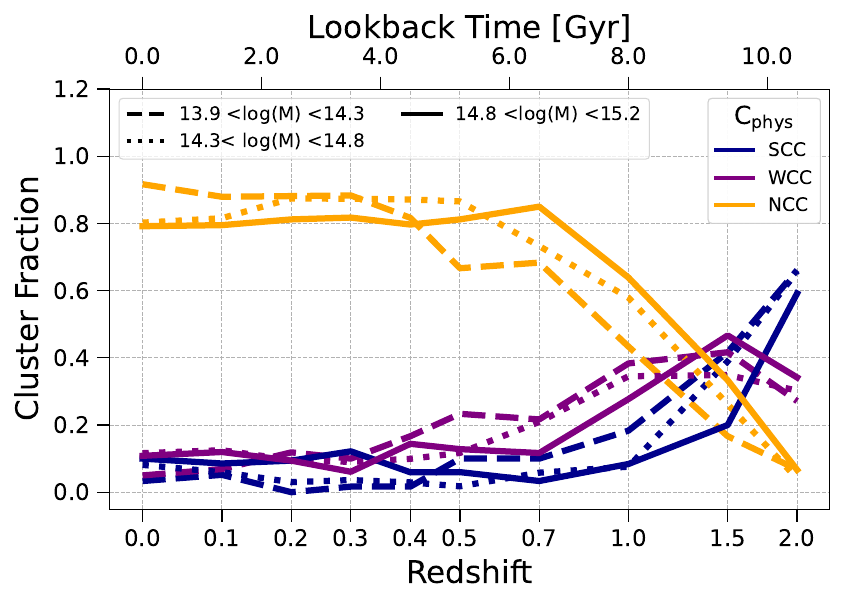}
    \includegraphics[width=0.42\textwidth]{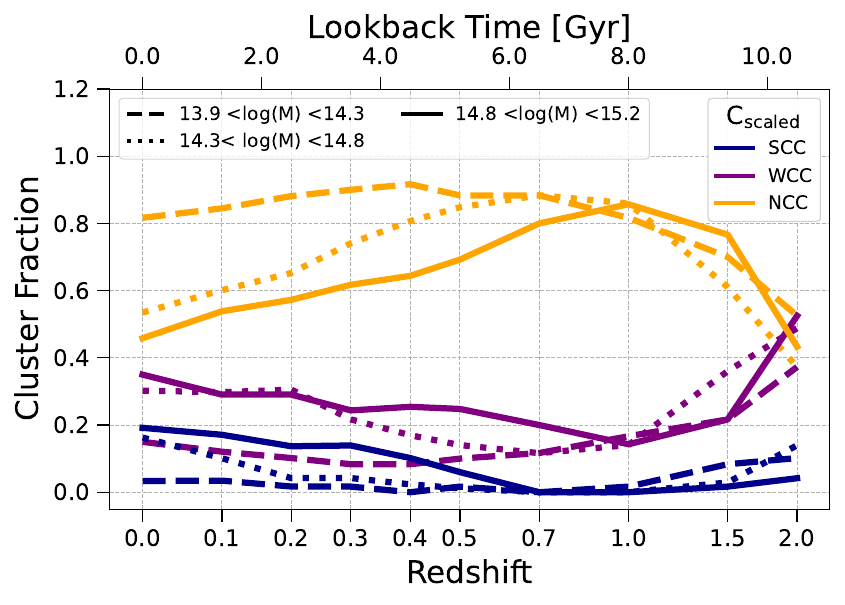}
    \caption{Fraction of SCC, WCC, and NCC clusters as a function of redshift. Analogous to Figure \ref{fig:fCCvsZ}, except that here the mass bins are defined at $z=0$, and the progenitors in each bin are tracked back to earlier redshifts.}
    \label{fig:fCCvsZ_classz0}
\end{figure*}

In the main text we study the redshift evolution of cool-core fractions and thermodynamical radial profiles. To do so we consider clusters at each redshift independently, and evaluate cool-core criteria given the current i.e. time-evolving properties of clusters. This is the information accessible in observations, making the results in the main text most comparable to data.

In this Appendix, we instead consider the redshift evolution of clusters by directly tracking their progenitors through time. 

Figure~\ref{fig:ProfilesVsRedshiftz0class} shows the entropy profiles for different cool-core classes but based on the $z=0$ cooling time of each cluster. We can see that the profiles for the three different cluster classes approach each other even in the core for increasing redshift. For $z = 2$, the order is even reversed: the non-cool-core cluster profiles have the smallest central entropy and the strong cool-core profile the largest, however the split between the three classes is small. At $z=1$ there are zero NCCs in our sample and most of the clusters are classified as SCCs (see Figure~\ref{fig:CCcritVsz}), thus the three lines represent the inner structure of SCCs. The NCCs at $z=0$ evolve from clusters with even lower central entropy at $z=1$ than SCCs, classified at $z=0$, have at that redshift.

Figure~\ref{fig:fCCvsZ_classz0} shows the fractions of strong cool-core (SCC), weak cool-core (WCC), and non-cool-core (NCC) clusters as a function of redshift, where the three halo mass bins are defined at $z=0$, and the progenitors of the clusters in each bin are tracked back. Each curve therefore shows the time evolution of a specific and fixed sub-sample of TNG-Cluster halos. Clearly, the SCC fractions are increasing towards earlier times for all CC metrics and all mass bins, while the NCC fractions are decreasing.

The CC fractions presented in Figure~\ref{fig:fCCvsZ}, show a less ambiguous picture. This is expected, as the mass within one fixed sub-sample is decreasing with redshift, and the decreasing mass was previously identified as the main driver of redshift evolution (Figure~\ref{fig:CCcritVsz}). With increasing redshift the central cooling times are strongly decreasing, such that at $z\leq 1$ and clusters are classified as SCC. In the upper left panel we can see, that the clusters that are high-mass clusters today are more rapidly transformed into NCCs than the less massive clusters today. This is expected as high-mass clusters undergo disruptive mergers at earlier times than low-mass clusters.

\section{Towards New Cool-Core Criteria and Sunyaev-Zeldovich Observables}\label{sec_appendix_newCCcrit}

{\renewcommand{\arraystretch}{1.3}
\begin{table*}
    \centering
    \caption{Summary of the 16th to 84th percentiles (left two columns), and 27th to 73rd percentiles (right two columns) of the distributions of the six ICM properties introduced in Sec.~\ref{ch:CCcritDef} and two properties of the SZ-signal (last two rows), that are well correlated with the central cooling time. See the text for a motivation of the percentile values and details of the two SZ quantities.}
  \begin{tabular}{lllll}
    \hline\hline
    Physical property &  SCC/WCC & WCC/NCC & SCC/WCC ($\bar{z}_{\rm form} = 0.52$) & WCC/NCC ($\bar{z}_{\rm form} = 0.52$)\\
    & threshold & threshold & threshold & threshold\\
    \hline
    $t_{\rm{cool},0}$  & 0.57\,Gyr& 7.7\,Gyr& 1.14\,Gyr& 5.36\,Gyr \\
    $K_0$ &33\,keV cm$^2$ & 237\,keV cm$^2$& 57\,keV cm$^2$ & 178\,keV cm$^2$ \\
    $n_{\rm{e},0}$&$ 8.1 \cdot 10^{-2}$\,cm$^{-3}$ & $0.6 \cdot 10^{-2}$\,cm$^{-3}$&$ 83.5 \cdot 10^{-2}$\,cm$^{-3}$ & $0.8 \cdot 10^{-2}$\,cm$^{-3}$\\
    $\alpha$ & 1 &  0.26& 0.7 &  0.34 \\
    $C_{\rm{phys}}$ & 0.1 & 0.017& 0.04 & 0.02\\
    $C_{\rm{scaled}}$ &0.48& 0.09&0.32& 0.1  \\
    \hline
    $\alpha_{\rm SZ}$ & 0.26 &  0.07& 0.18 &  0.08 \\
    $C_{\rm{phys, SZ}}$ & 0.024 & 0.013& 0.02 & 0.014\\
    \hline\hline
  \end{tabular}
  
  \label{tab:CCcutsNew}
\end{table*}}

In this Appendix, we discuss a new perspective on the distributions of core properties, as presented in Figure~\ref{fig:HistCCcrit}, and cool-core classification. Furthermore, we introduce several new possible CC criteria which depend on the Sunyaev-Zeldovich signal of clusters. 
Figure~\ref{fig:HistCCcrit} clearly shows that the distributions of CC criteria are unimodal. Unimodal distributions of central ICM properties are also found in observations \citep{andrade-santos2017, croston2008, pratt2010} and other simulations \citep{barnes2018, kay2007}. Given that the distributions appear unimodal, the observationally motivated thresholds for classifying CCs do not align with distinct features in the CC criteria distributions. Rather, we find that SCCs and NCCs represent the extremes of these distributions. The width of the wings of a distribution can be quantified by the percentiles corresponding to the $1\sigma$ interval. The respective values of each of the six distributions are stated in columns 2 and 3 of Tab.~\ref{tab:CCcutsNew}. By coincidence the 84 percentile of the distribution of the central cooling time corresponds to the NCC threshold of 7.7\,Gyr.

We propose that these $1\sigma$ intervals can not only be used to quantify the distributions of CC criteria, but also as a new definition of the core status of a cluster. In that case all clusters with $t_{\rm cool, 0} > 7.7$\,Gyr are classified as NCCs and clusters with $t_{\rm cool, 0} < 0.57$\,Gyr as SCCs. This approach fixes the SCC and NCC fraction to $\sim 16\%$ at $z=0$ for a sample of clusters with our mass distribution. Of course, to fully make this classification scheme useful and broadly applicable, the thresholds presented in Tab.~\ref{tab:CCcutsNew} can be refined by more carefully selecting the underlying cluster sample used to determine the percentile values.

Is it possible to derive an ideal percentile to select clusters using theoretical arguments? Of the six CC thresholds presented in Tab.~\ref{tab:CCcuts}, only the cooling time for NCCs is theoretically motivated. The other thresholds are deduced from distributions of observed samples, many of which originated in older, biased samples. 

The NCC cooling time threshold is based on the assumption that a cluster formed at $z=1$ should have a lower central cooling time than the lookback time at $z=1$. This condition ensures that the cluster can cool rapidly enough to develop a cool core. By considering the median formation redshift of the clusters in our sample, we can refine this argument. At the median formation redshift of $\bar{z}_{\rm form} = 0.52$ the lookback time is $t_{\rm L} = 5.36$\,Gyr. This particular value corresponds to the 73th percentile of the distribution of the central cooling times. We can use this width of $\sim 27\%$ to classify SCCs and NCCs. The respective values for the thresholds are stated in column 4 and 5 in Tab.~\ref{tab:CCcutsNew}. 

\begin{figure*}
    \centering
	\includegraphics[width=\textwidth]{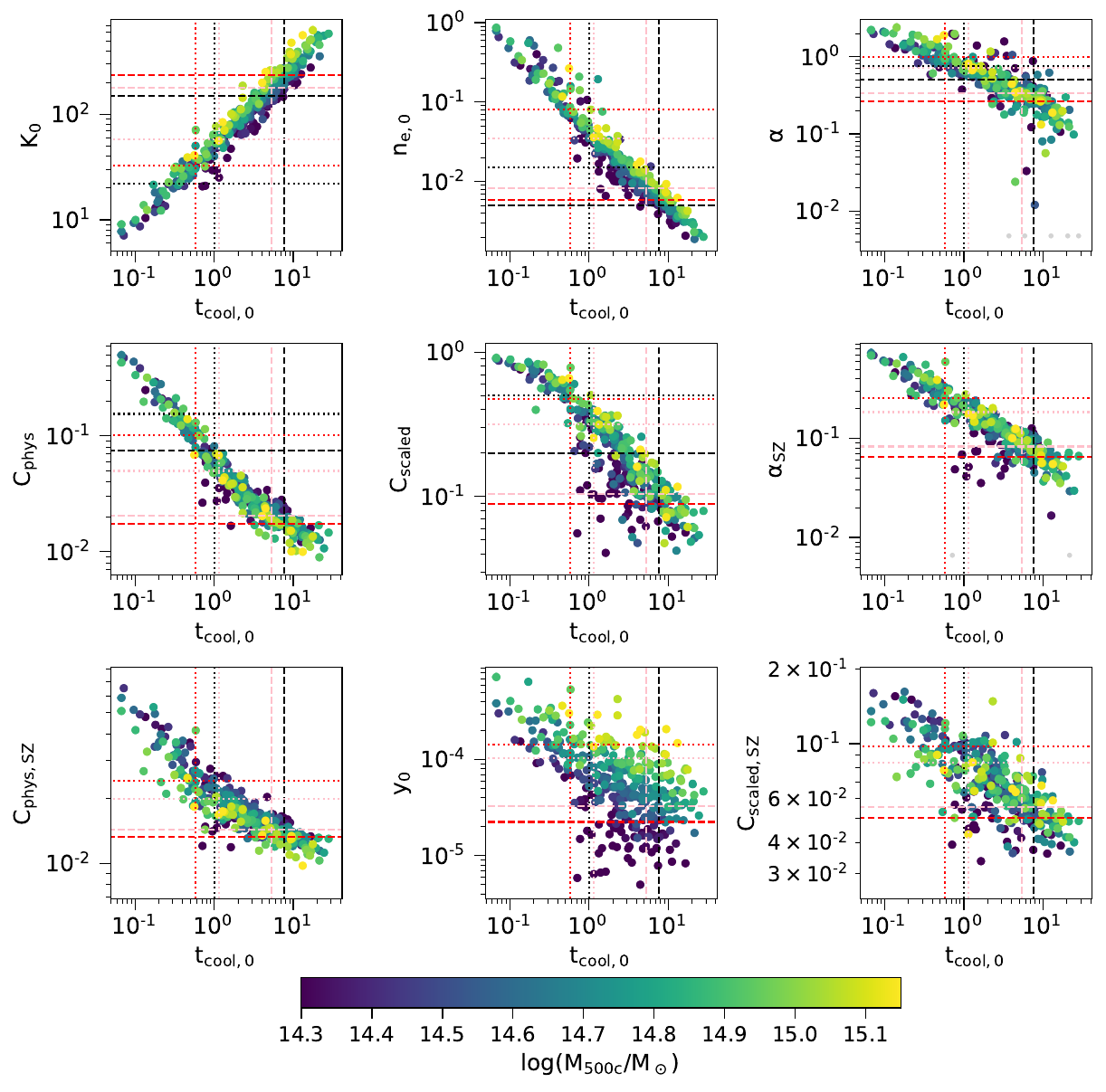}
    \caption{Correlation of ICM properties with the central cooling time for the whole TNG-Cluster sample at $z=0$. We show the correlation with the five other properties introduced in Sec.~\ref{ch:CCcritDef} and four new properties deduced from the Sunyaev-Zeldovich signal. The mass of each cluster is specified by the color of the dots. Black lines indicate the SCC/WCC thresholds (dotted line) and WCC/NCC thresholds (dashed line) taken from the literature (see Tab.~\ref{tab:CCcuts}). The red lines indicate the $1\sigma$ thresholds (column 2 and 3 in Tab.~\ref{tab:CCcutsNew}), whereas the pink lines show the thresholds stated in column 4 and 5 of Tab.~\ref{tab:CCcutsNew}. (See the text for a motivation of these thresholds.) The grey dots indicate clusters with a negative cuspiness. The five CC metrics considered in this paper as well as two metrics computed from the SZ signal are well correlated with the central cooling time.}
    \label{fig:CCcritCorrel}
\end{figure*}

To translate this to other CC criteria, we first check whether other ICM central properties are well correlated with the central cooling time. Figure~\ref{fig:CCcritCorrel} shows the correlations of the central cooling time with the other five core properties (first five panels). All properties are well correlated with the central cooling time.

The other four panels show CC criteria candidates computed from the Sunyaev-Zeldovich signal.\footnote{For the thermal SZ signal we compute Compton y-parameter maps by integrating the line-of-sight electron pressure, following \textcolor{blue}{Nelson et al. (submitted)}.} This is an idea which has not yet been applied in observations. These four panels show four quantities derived from spatially resolved SZ maps: $\alpha_{\rm SZ}$ (following the same definition of the X-ray cuspiness parameter), $C_{\rm phys, SZ}$ (likewise for $C_{\rm phys} $), $y_0$ (the Compton y-parameter within $r< 0.012 \rvir$), $C_{\rm scaled, SZ}$ (the same definition as $C_{\rm scaled} $). 

We find that both $y_0$ and $C_{\rm scaled, SZ}$ are only weakly correlated with the central cooling time, making them less suitable for core state classification. In contrast, $\alpha_{\rm SZ}$ and $C_{\rm phys, SZ}$ are well correlated with the central cooling time, motivating core status classification using one of these parameters. We provide the threshold values for these two parameters in Tab.~\ref{tab:CCcutsNew}.

In future work we suggest that new, optimal CC criteria can be developed by combining multi-wavelength (mock) cluster observables combined with the ground truth from TNG-Cluster. Given that different tracers, from the X-ray, to SZ, to optical, to radio, all contain non-degenerate information content, they can jointly constrain the central cooling time and thus cool-core state of a cluster better than any one observable alone.

\end{appendix}

\end{document}